\pdfoutput=1
\documentclass[12pt]{article} 
\usepackage{alphabeta}
\usepackage{amsmath}
\usepackage{graphicx}
\usepackage{subcaption}
\usepackage{titlesec}

\usepackage[showcomments]{ajtex}
\DeclareRobustCommand{\DIEP}{\ensuremath{%
    \mathchoice{\includegraphics[height=2ex]{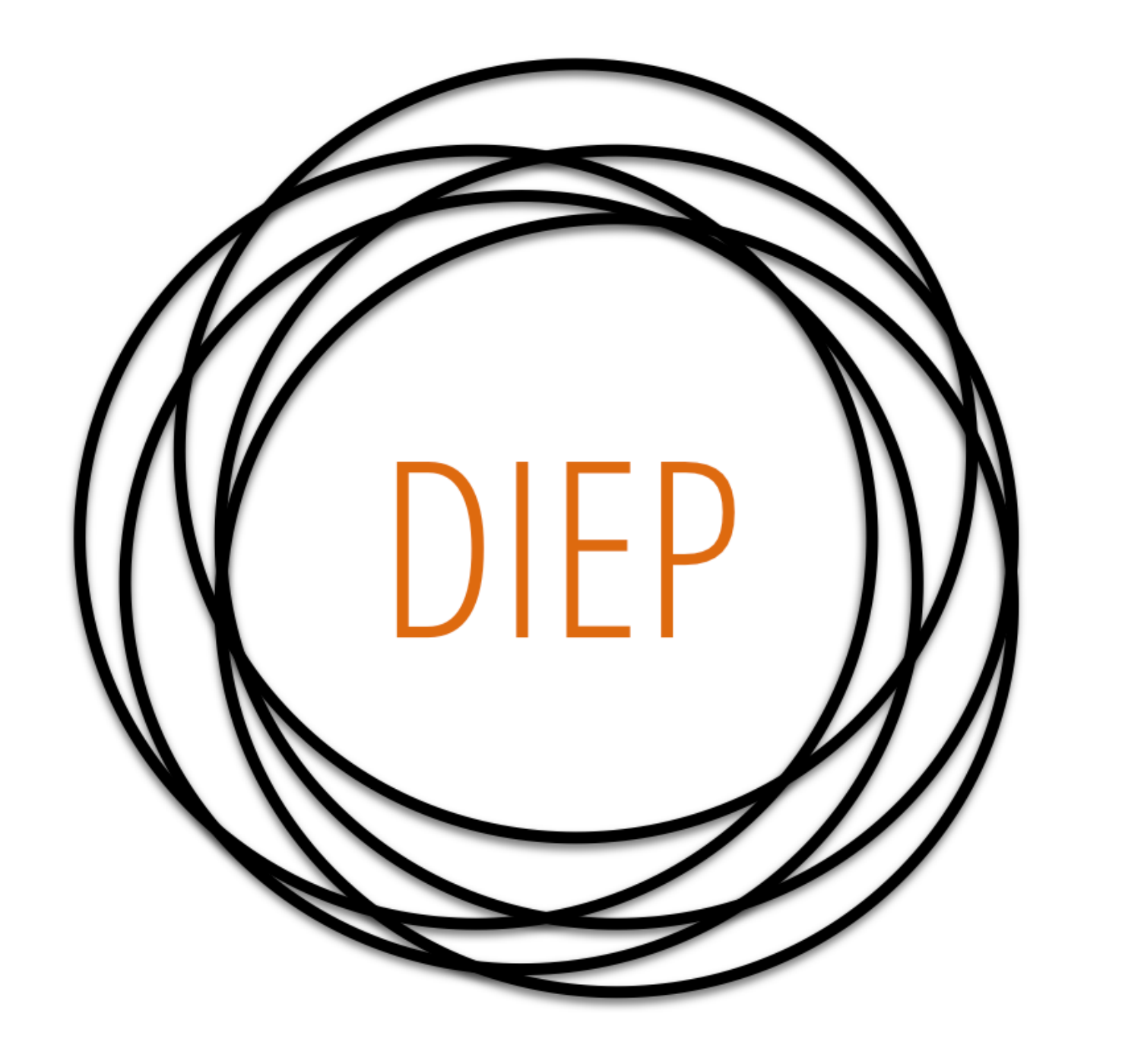}}
    {\includegraphics[height=2ex]{DIEPs.pdf}}
    {\includegraphics[height=1.5ex]{DIEPs.pdf}}
    {\includegraphics[height=1ex]{DIEPs.pdf}}
  }}

\title{Holographic duals of the $\mathcal{N}=1$* gauge theory}

\author[a,b,\DIEP]{Jay Armas}\email{j.armas@uva.nl}

\author[a,b]{Giorgos Batzios}\email{g.batzios@uva.nl}

\author[a,b]{Jan Pieter van der Schaar}\email{J.P.vanderSchaar@uva.nl}

\affiliation[a]{Institute for Theoretical Physics, University of Amsterdam, 1090
  GL Amsterdam, The Netherlands}

\affiliation[b]{Delta Institute for Theoretical Physics, Science Park 904, PO Box 94485, 1090 GL Amsterdam, The Netherlands}

\affiliation[\DIEP]{Dutch Institute for Emergent Phenomena, 1090 GL Amsterdam, The Netherlands}

\date{6th of December, 2022}

\abstract{We use the long-wavelength effective theory of black branes (blackfold approach) to perturbatively construct holographic duals of the vacua of the $\mathcal{N}=1$* supersymmetric gauge theory. Employing the mechanism of Polchinski and Strassler, we consider wrapped black five-brane probes with D3-brane charge moving in the perturbative supergravity backgrounds corresponding to the high and low temperature phases of the gauge theory. Our approach recovers the results for the brane potentials and equilibrium configurations known in the literature in the extremal limit, while away from extremality we find metastable black D3-NS5 configurations with horizon topology $\mathbb{R}^3\times \mathbb{S}^2\times\mathbb{S}^3$ in certain regimes of parameter space, which cloak potential brane singularities. We uncover novel features of the phase diagram of the $\mathcal{N}=1$* gauge theory in different ensembles and provide further evidence for the appearance of metastable states in holographic backgrounds dual to confining gauge theories.}

\DeclareMathAlphabet{\mathdutchcal}{U}{dutchcal}{m}{n}
\SetMathAlphabet{\mathdutchcal}{bold}{U}{dutchcal}{b}{n}
\DeclareMathAlphabet{\mathdutchbcal}{U}{dutchcal}{b}{n}

\usepackage[bbgreekl]{mathbbol}

\setlist[itemize]{itemsep=0pt,topsep=0pt}

\begin{document}

\maketitle

\section{Introduction}

The AdS/CFT correspondence \cite{maldacena1999large,2000} postulates the equivalence between type IIB string theory on $AdS_{5}\times S^{5}$ and 4-dimensional $\mathcal{N}=4$ Super Yang-Mills (SYM) theory with gauge group $SU(N)$ and effective coupling $λ=Ng_{Y M}^{2}$. In the large N and large $λ$ limit solutions of classical gravity in the bulk correspond to states of the field theory living on the boundary. A remarkable property of this conjecture derived from string theory is that it can account for non-conformal theories with (partially or completely) broken supersymmetry \cite{witten1998anti1,witten1998anti2}. Over the past many years, a considerable amount of work has been done in the direction of mapping calculations from a suitable strongly coupled field theory to a classical theory of gravity. Of particular interest are field theories that share certain features with Quantum Chromodynamics (QCD) for which an in-depth study can lead to a better understanding of how to resolve the long-standing problem of confinement in gauge theories.

A concrete example of such a QCD-like theory is the $\mathcal{N}=1$* supersymmetric theory which is obtained as a deformation of $\mathcal{N}=4$ SYM by adding mass terms for the 3 chiral multiplets. It possesses a rich phase space, with confining and deconfining phases interrelated by phase transitions \cite{Donagi_1996}. Importantly, it has a well defined ultraviolet fixed point in contrast with other known $\mathcal{N}=1$ theories, and it can be studied via lattice simulations as well as holographic methods. On the gravity side, Polchinski and Strassler \cite{Polchinski:2000uf} proposed that the vacua of the gauge theory correspond in the bulk to five-branes with D3-brane charge wrapping an equator of the $S^{5}$, in an asymptotically $AdS_{5}\times S^{5}$ spacetime. The addition of mass deformations to $\mathcal{N}=4$ means, in the 10-dimensional type IIB language, turning on non-normalizable modes of the 3-forms on the $S^{5}$. Their 7-form duals couple to the D3-branes which by Myers' effect \cite{Dielectic-branes} polarize into five-branes sitting at a non-zero Anti-de Sitter (AdS) radius. We will see that this picture for the holographic dual of the $\mathcal{N}=1$* detailed by Polchinski and Strassler \cite{Polchinski:2000uf} can indeed be described directly in the supergravity regime.

The construction of the backreacted supergravity solution dual to the vacua of the $\mathcal{N}=1$* theory has remained a daunting task since the probe analysis of \cite{Polchinski:2000uf} in the low temperature phase of the theory, below the expected Hawking-Page transition in the canonical ensemble. In this regime, the GPPZ flow \cite{2000_GPPZ} obtained in the context of 5-dimensional gauged supergravity, and its uplift to a 10-dimensional solution of type IIB supergravity on $AdS_{5}\times S^{5}$, is a promising candidate for a fully backreacted supergravity solution that holographically describes a class of vacua of the gauge theory. However, the 5-dimensional GPPZ solution, as well as the corresponding 10-dimensional uplift, contains a singularity \cite{Petrini:2018pjk,2018_uplift_GPPZ}. The origin of this singularity has been studied in detail \cite{Petrini:2018pjk,2019} and has led to the debate of whether the GPPZ solution actually corresponds to any of the supersymmetric vacua of the $\mathcal{N}=1$* theory. Recently, in \cite{2019} it was shown that the near-singularity structure of the uplifted GPPZ solution admits an interpretation in terms of smeared five-branes along a compact direction of spacetime. Though the existence of five-branes is a neccessary ingredient, it is still yet to be shown that the GPPZ flow is the holographic dual to the Coloumb vacua of the gauge theory as described by Polchinski and Strassler \cite{Petrini:2018pjk,2018_uplift_GPPZ,2019,Bena_2019}.\footnote{In section \ref{3.0} we describe in some detail the supersymmetric vacua of the $\mathcal{N}=1$* theory following \cite{Polchinski:2000uf}. }

Progress has also been made in the high-temperature phase of the theory, above the expected Hawking-Page transition. In this regime, Minahan and Freedman \cite{Friedman} argued that five-branes are not essential since infrared singularities arising from the 3-form perturbation are hidden behind the horizon, and constructed perturbatively the supergravity solution involving an asymptotically $AdS_{5}\times S^{5}$  black hole with spatial horizon topology $\mathbb{R}^3\times \mathbb{S}^5$, at second order in the mass perturbation $m$. This solution describes a high-temperature deconfined vacuum of the gauge theory. Recently, a fully backreacted supergravity solution of this high-temperature phase was found using numerical techniques \cite{Bena_2019}, whose properties are in accordance with the perturbative results of \cite{Friedman}. However, Ref.\cite{Bena_2019} already pointed out that this is not necessarily the only deconfined vacuum of the $\mathcal{N}=1$* theory in the high-temperature phase. In fact, their numerical analysis hinted at the existence of a topological phase transition from horizon topology $\mathbb{R}^3\times \mathbb{S}^5$ to $\mathbb{R}^3\times \mathbb{S}^2\times \mathbb{S}^3$ mediated by a Gregory-Laflamme-type of instability. This high-temperature phase with horizon topology $\mathbb{R}^3\times \mathbb{S}^2\times \mathbb{S}^3$ admits an interpretation in terms of polarised D3-branes for which there is no explicit supergravity construction.

The main focus of this paper is in finding holographic duals with $\mathbb{R}^3\times \mathbb{S}^2\times \mathbb{S}^3$ topology. Such supergravity solutions describe massive vacua of the $\mathcal{N}=1$* theory, and verifying their existence remains an important open problem as it is key to uncovering various phase transitions in the theory. Our approach, contrary to the DBI probe analysis of \cite{Polchinski:2000uf} at zero temperature and the analysis of \cite{Friedman} at high temperature, consists in tackling the problem of finding holographic duals directly in the 10-dimensional supergravity regime. This can be accomplished by employing a long-wavelength effective theory, known as the blackfold approach \cite{Emparan_2010,Blackfolds_string_theory}, in order to perturbatively construct the supergravity solution. This method has been used in various holographic contexts to analyze anti-D3 branes at the tip of the Klebanov-Strassler throat \cite{PhysRevLett.122.181601} and anti-M2 branes in the Cvetic-Gibbons-Lu-Pope \cite{m2-m52019} background, as well as to construct thermal giant gravitons \cite{2012a, Armas:2013ota}, thermalised Wilson loops \cite{2012b, Armas:2014nea} in AdS/CFT and find new black holes in AdS spacetime \cite{Armas:2010hz, Armas:2015qsv}. We provide various details about the method in section \ref{2.0}. Our goal in this paper is to apply this same methodology in which D3-NS5 branes are wrapped around an $S^2$ of mass deformed AdS in order to perturbatively construct supergravity solutions with $\mathbb{R}^3\times \mathbb{S}^2\times \mathbb{S}^3$ topology at low- and high-temperatures.

The analysis of Polchinski and Strassler (PS) revealed that the supersymmetric vacua of $\mathcal{N}=1$* gauge theory are metastable \cite{Polchinski:2000uf}. The same metastability properties were observed by Freedman and Minahan (FM) in the high-temperature phase \cite{Friedman}. Our method allows to perturbatively construct these metastable states directly in the supergravity regime. By wrapping extremal D3-NS5 branes in mass deformed AdS or in mass deformed black D3-brane backgrounds, we show that such solutions are indeed metastable and consistent with \cite{Polchinski:2000uf, Friedman}. Yet, similarly to the case of D3-NS5 branes at the tip of Klebanov-Strassler throat \cite{PhysRevLett.122.181601}, one may wonder about the fate of infrared singularities due to the presence of brane sources. Bypassing this issue, we in addition consider wrapping non-extremal (black) D3-NS5 branes in the same backgrounds in order to cloak brane singularities. This allows us to show the existence of metastable solutions with spatial horizon topology $\mathbb{R}^3\times \mathbb{S}^2\times \mathbb{S}^3$, which are continuously connected to the PS supersymmetric vacua at low temperatures and to the FM deconfined vacua at high-temperatures. Furthermore, metastability is lost at a critical value of the global entropy in the system, or at a critical value of the temperature in the high-temperature phase, where two classes of solutions (an unstable and a metastable) meet. This picture is qualitatively similar to that of D3-NS5 branes in Klebanov-Strassler \cite{PhysRevLett.122.181601}.

We study the phase diagram of the $\mathcal{N}=1$* gauge theory in the constant entropy (or isentropic) and canonical ensembles, including in our analysis known analytic and numerical solutions as well as the two perturbative solutions found in this work. In particular, we show that the perturbative solution we construct that is continuously connected to the PS vacua can dominate the constant entropy ensemble in a specific regime of parameters but is always subdominant in the canonical ensemble. In the canonical ensemble this solution is approached at high-temperatures, similarly to the GPPZ solution as argued in \cite{Bena_2019}. In turn, the second solution we construct that is continuously connected to the FM solution is a black hole solution with two disconnected horizons which are generically coexisting at different temperatures. We show that, to leading order in the perturbative scheme we employ, there is no choice of parameters for which the two horizons are in thermal equilibrium. As such, these solutions do not have a role to play in the phase diagrams of these ensembles. Our work thus sheds light on the various phases of $\mathcal{N}=1$* gauge theory and potential phase transitions. 

This paper is organised as follows. In section \ref{2.0} we introduce the basics of the blackfold effective theory applied to the specific supergravity context that we are interested in. In section \ref{3.0} we initiate the study of the Polchinski-Strassler construction by embedding five-brane probes in a perturbed $AdS_{5}\times S^{5}$ geometry in the presence of fluxes, describing the low temperature phase of the theory. In the extremal limit we find that the PS vacua are consistent with the blackfold equations. In the non-extremal regime, we perturbatively find stationary solutions to the blackfold equations with horizon topology $\mathbb{R}^3\times \mathbb{S}^2\times \mathbb{S}^3$ which are continuously connected with the PS vacua. These solutions exist up to a critical value of the global entropy of the system. At that point, a transition occurs in which the metastable state merges with an unstable configuration. In section \ref{4.0} we consider five-branes on top of the high-temperature deconfined vacuum of the theory. We therefore consider non-extremal probes on the background of the perturbed black hole geometry of ref. \cite{Friedman}, focusing on the physics far away from the background horizon. In the extremal limit there is a maximum value for the background temperature that permits the formation of a metastable state \cite{Friedman} and we find that such a critical value is also present in the non-extremal regime. For temperatures lower than this critical value, our results are qualitatively similar to the analysis of the low temperature phase and we recover a metastable black D3-NS5 configuration. As expected, thermal effects associated with the internal processes of the bound state dominate the free energy of the probes for sufficiently low, though not arbitrarily small, background temperatures. We conclude in section \ref{5.0} with a summary of the results and a discussion of some of the open questions and future directions.


\section {Elements of the effective theory}\label{2.0}
In this section we review and partly extend the blackfold approach applied to D3-NS5 branes \cite{Emparan_2010, Blackfolds_string_theory, 2016, PhysRevLett.122.181601}. This approach can be viewed as a generalised fluid/gravity correspondence in which one maps the long-wavelength deformations of a (black) brane into a hydrodynamic system restricted to a higher-dimensional surface of spacetime - the blackfold. The fluid moving on this submanifold obeys constraint equations that correspond to necessary (sufficient in all known cases) conditions for the existence of a regular backreacted solution \cite{Camps_2012,Niarchos_2016,Nguyen:2021srl} which locally, near the horizon is approximated by a uniform, flat p-brane solution (in our case the D3-NS5 or D3-D5 bound states) and asymptotically approaches a given background solution \cite{2016,m2-m52019} - in this particular case it approaches mass deformed $AdS_{5}\times S^{5}$. Below we discuss the geometry of black D3-NS5 branes, blackfold constraint equations and thermodynamics of putative solutions.

\subsection{Near-horizon solution}
Our goal is to construct perturbative solutions of type IIB supergravity that approximate the mechanism of polarized branes of Polchinski and Strassler for the holographic dual of the $\mathcal{N}=1$* theory. We are interested in a sector of solutions which can be obtained in a long-wavelength (small derivative) expansion of the supergravity equations. One can then attempt to build an interpolating solution where the near-horizon geometry and fluxes of a wrapped five-brane are matched with the perturbative backgrounds found in \cite{Polchinski:2000uf,Friedman}. 

In order to realise this perturbative scheme within the context of the blackfold approach, we denote collectively as $r_{b}$ the characteristic scales of the near-horizon solution, while $\mathcal{R}$ and $\mathcal{L}$ denote the scales associated with the worldvolume and background fields, respectively. We are interested in the region of parameter space for which
\begin{equation}\label{hierarchy}
    r_{b}\ll min(\mathcal{R},\mathcal{L})~~.
\end{equation}
 The existence of the small parameters $\frac{r_{b}}{\mathcal{R}}$ enables a matched asymptotic expansion in which the near-horizon solution valid at distances $r\ll \mathcal{R}$ is matched to a far-zone (or asymptotic) background solution valid at large distances $r\gg r_{b}$ in the overlap region $r_{b}\ll r \ll \mathcal{R}$. On the other hand, the small parameters $\frac{r_{b}}{\mathcal{L}}$ ensure that the near-horizon solution of a wrapped bound state of branes can be approximated at leading order by the solution of the bound state in flat space. Up to now a general and systematic treatment of matched asymptotic expansions of this kind is an open problem. Thus, the existence of a regular deformed black brane solution remains a case-by-case study. 
 
 In the particular case we consider here, the corresponding non-extremal, asymptotically flat D3-NS5 black brane solution of type IIB supergravity obtained by S-duality of the respective D3-D5 solution, is given in the Einstein frame by
\begin{equation}\label{branegeometry}
    ds^2=(HD)^{-\frac{1}{4}}\left(-fdt^2+D((dx^{1})^2+(dx^{2}))^2+\sum_{i=3}^{5}(dx^{i})^2\right)+H^{\frac{3}{4}}D^{-\frac{1}{4}}(f^{-1}dr^{2}+r^{2}dΩ^{2}_{3})~~,
\end{equation}
accompanied by the fields 
\begin{gather}\label{branefluxes}
C_{2}=-\tan{θ}(Η^{-1}D-1)dx^{1}\wedge dx^{2} \quad, \quad B_{2}=-r_{0}^2\sinh2{\alpha}\cos{\theta}(\phi)\sin^2{\psi}\sin{\omega}d\psi\wedge d\omega~~, \nonumber \\
C_{4}=(Η^{-1}-1)\coth{α}\sin{θ}dt\wedge dx^{3}\wedge dx^{4}\wedge dx^{5}+B_{2}\wedge C_{2}\left(1+\frac{r^2}{r_{0}^2\sinh^2{\alpha}\cos^2{\theta}}\right)~~, \\
e^{2Φ}=HD^{-1}~~, \nonumber
\end{gather}
where
\begin{equation}
   f(r)=1-\left (\frac{r_{0}}{r}\right )^{2}\quad ,\quad H(r)=1+\left (\frac{r_{0}}{r}\right )^{2}\sinh^{2}{α}\quad,\quad D=(\sin ^{2}{θ}Η^{-1}+\cos^{2}{θ})^{-1}~~.
\end{equation}
Here and in what follows we set the asymptotic value of the dilaton $Φ_{\infty}$ equal to zero. We have also expressed the $S^{3}$ part of the metric using $d\Omega_{3}^2=d\psi^2+\sin^2{\psi}(d\omega^2+\sin^2{\omega}d\phi^2)$. The parameters $r_{0},α,θ$ characterize the non-extremal solution and parametrize the thermodynamic quantities of the brane. In the blackfold approach, they are promoted to slowly varying fields with support on the six-dimensional worldvolume of the brane spanned by the coordinates $σ^{b}$, $b=0,1...5$. The extremal solution is recovered by taking the limit 
\begin{equation}\label{extremalilimit}
    r_{0}\rightarrow 0 \quad,\quad α\rightarrow \infty~~,
\end{equation}
in such a way that the combination $r_{0}^2e^{2α}$ remains fixed. As soon as $r_{0}$ and $θ$ become non-zero, the SO(5,1) symmetry of the worldvolume is broken by the introduction of an entropy flow and the presence of D3 charge. The associated Goldstone modes are encoded in the timelike velocity vector field $u^{a}$ and the spacelike orthonormal vector fields $v^{a},w^{b}$. The geometry \eqref{branegeometry} sources the gravitational field in the asymptotic region of spacetime given by the Brown-York stress-energy tensor 
\begin{equation}\label{stress1}
    \mathcal{T}^{a b}=T^{a b}δ^{4}(x^{μ}-X^{μ})~~,
\end{equation}
where
\begin{equation}\label{stress2}
    T^{a b}=\mathcal{C}\left( r_{0}^2(u^{a}u^{b} -\frac{1}{2}γ^{a b}) - r_{0}^{2}(\sinh{α})^{2}(\sin{θ})^{2}(γ^{a b} - v^{a}v^{b} - w^{a}w^{b}) - r_{0}^{2}(\sinh{α})^{2}(\cos{θ})^{2}γ^{a b} \right).
\end{equation}
In \eqref{stress2} we have promoted the flat worldvolume metric $η_{a b}$ to a slowly varying induced metric,
\begin{equation}
      γ_{a b}=\partial_{a}X^{μ}\partial_{b}X^{ν}g_{μ ν}~~,
\end{equation}
 to allow for extrinsic deformations of the brane induced by its embedding $X^{μ}(σ^{a})$ in the ambient 10-dimensional background geometry $g_{μ ν}$ with coordinates $x^{μ}$. The delta function in equation \eqref{stress1} expresses the localization of the degrees of freedom on the brane. We have defined $\mathcal{C}=\frac{Ω_{3}}{8\pi G_{10}}$ where $G_{10}$ is Newton's constant in 10 dimensions.
 
 From the point of view of supergravity, D/NS branes are extended charged objects which can be coupled to the gauge fields of the theory. We may assign to a (stack of) brane(s) a current for each gauge field under which the brane is charged, sourcing the charge in the background spacetime. Upon inspection of the fluxes of the solution \eqref{branefluxes} we see that the D3-NS5 black brane carries the asymptotic higher-form currents
\begin{equation}\label{highercurrents}
    J_{2}\quad,\quad J_{4}\quad,\quad j_{6} \quad .
\end{equation}
Here, $J_{2}$ and $J_{4}$ are electric currents sourcing the gauge fields $C_{2}$, $C_{4}$ while $j_{6}$ is a magnetic current corresponding to the 3-form $H_{3}$ of the solution. These currents are formally computed using the sourced equations of motion of supergravity, in a small $\frac{r_{0}}{r}$ expansion. When the background into which the brane is embedded has a dilaton that depends on the brane coordinates transverse to the worldvolume $y$, i.e. $Φ=Φ(y)$, then the above currents have to be rescaled appropriately. This can be readily seen from the requirement that a constant shift of the dilaton is a symmetry of the worldvolume action (see eq.~\eqref{action} below). Taking this into account, the currents in the Einstein frame read (omitting the delta functions)
\begin{gather}
j_{6}=\mathcal{C}e^{-\frac{φ}{2}}r_{0}^{2}\sinh{α}\cosh{α}\cos{θ}(\ast 1) ~~, \\ 
J_{4}=\mathcal{C}r_{0}^{2}\sinh{α}\cosh{α}\sin{θ}\ast (v\wedge w)~~,\\ \label{currents}
J_{2}=\mathcal{C}e^{-\frac{φ}{2}}r_{0}^{2}(\sinh{α})^{2}\sin{θ}\cos{θ}v\wedge w ~~,
\end{gather}
where $φ=Φ(X^{μ}(σ,y))$ is the pullback of the dilaton onto the worldvolume. In the above 3 expressions the Hodge dual is taken with respect to the induced metric $γ_{a b}$. Moreover, there is a running dilaton on the worldvolume of the brane to which we associate the current $\mathcal{J}_{Φ}$ with the form
\begin{equation}
    \mathcal{J}_{Φ}=j_{Φ}δ^{4}(x^{μ}-X^{μ}) ~~.
\end{equation}
Using the sourced equation of motion for a time-independent dilaton we compute
\begin{equation}
  j_{Φ}=-\frac{\mathcal{C}}{2}\cos^2{θ}r_{0}^2\sinh^2{α} ~~.
\end{equation}
For our purposes we also need the energy density $ε$, local temperature $\mathcal{T}$, and local entropy density $s_{0}$ of the brane 
\begin{equation}
  ε=\mathcal{C}r_{0}^{2}(\frac{3}{2}+\sinh^2{α})\quad ,\quad   s_{0}=2π\mathcal{C}r_{0}^3\cosh{α} \quad ,\quad  \mathcal{T}=\frac{1}{2πr_{0}\cosh{α}}~~. \label{thermodynamicdata}
\end{equation}\\
Before closing this section, we note that the effective blackfold description of the D3-NS5 system that we consider here is straightforwardly extended to the D3-D5 system since its thermodynamic quantities are unaffected by S-duality.

\subsection{Blackfold equations}
Solving the supergravity equations within this perturbative scheme requires satisfying constraint equations, which determine the blackfold effective theory on the brane. These equations can be derived in full generality by coupling the supergravity equations of motion to generic brane sources \cite{2016}. Within the context of the matched asymptotic expansion envisioned here, this derivation is valid in the far-region. 

Given the sources of the D3-NS5 bound state, the general leading order blackfold equations, assuming a vanishing Ramond-Ramond (RR) scalar field, are given by \cite{2016}
\begin{align}\label{blackfoldequations}
    \nabla_{μ}Τ^{μ ν}&=\frac{1}{2} \tilde{F}^{ν μ_{1} μ_{2}}_{3}J_{2 μ_{1}μ_{2}}+ \frac{e^{-Φ}}{6!}H^{ν μ_{1}...μ_{6}}_{7}j_{6 μ_{1}...μ_{6}}+\frac{1}{4!}\tilde{F}^{ν μ_{1}..μ_{4}}_{5}J_{4 μ_{1}..μ_{4}} + j_{Φ}\partial^{ν}Φ \\& \nonumber \quad \quad \qquad \qquad +\frac{1}{4!}\left(3H_{3}^{νμ_1μ_2}C_{2}^{μ_{3}μ_{4}}\right)J_{4 μ_{1}..μ_{4}}~~.
\end{align}
Here, the RR field strengths are defined as $\tilde{F}_{q+2}=F_{q+2}-H_{3}\wedge C_{q-1}$, $F_{q+2}=dC_{q+1}$ for $q=1,3$, while the 7-form is given by $H_{7}=\star H_{3}$ where the Hodge duality is written with respect to the background geometry $g_{μ ν}$. The equations governing the evolution of the currents are obtained as Bianchi identities of the type IIB supergravity equations, leading to 
\begin{gather}
  d\star J_{2}+H_{3}\wedge \star J_{4}+\star j_{6}\wedge \star \tilde{F}_{5}=0~~, \nonumber\\ 
 d\star J_{4}-\star j_{6}\wedge F_{3}=0~~, \label{conservation1} \\ 
  d\star j_{6}=0~~.  \nonumber
\end{gather}
From the point of view of an asymptotic observer in the far-zone region, the continuity equations \eqref{blackfoldequations} supplemented with \eqref{conservation1} constitute the equations of motion for the brane in a given asymptotic background spacetime in the presence of fluxes, and represent a set of forced hydrodynamical equations for a fluid moving on a dynamical surface embedded in spacetime.

We first focus on the relevant conservation equations for the currents. From the last equation in \eqref{conservation1} we deduce the conservation of the charge 
\begin{equation}\label{definitionQ5}
     Q_{5}=\int \ast j_{6}~~,
 \end{equation}\\
 where the Hodge dual is defined with respect to the worldvolume metric. Since $F_{3}=dC_{2}$ the second equation in \eqref{conservation1} is equivalent to
 \begin{equation}
     d\left(\star J_{4}-(\star j_{6}\wedge C_{2})\right)=0~~,
 \end{equation}
 implying the conservation of the Page 3-charge \\
 \begin{equation}\label{definitionQ3}
      \tilde{Q}_{3}=\int \ast (J_{4} + \star(\star j_{6}\wedge C_{2}) )~~.
 \end{equation}
  The quantization conditions \cite{https://doi.org/10.48550/arxiv.hep-th/0006117} for the charges \eqref{definitionQ5}, \eqref{definitionQ3} imply that they must be proportional to the total number of NS5 and D3 branes respectively, i.e.
 \begin{equation}\label{quantizationconditions}
        Q_{5}\sim N_{5} \quad,\quad  \tilde{Q}_{3}\sim N_{3}~~.
 \end{equation}
The proportionality constants will be fixed in section \ref{3.2} and are related to the tension of each of the coincident NS5 and D3 branes, respectively.

\subsection{Equilibrium partition function}\label{Effectiveaction}
The long-wavelength effective theory of black branes maps the thermal excitations of the brane to an effective relativistic fluid on a dynamical surface. To write an action for a relativistic fluid in a general supergravity setting is a non-trivial task. When restricting to stationary configurations, however, the situation becomes tractable. We thus assume the existence of a background killing vector $k^{\mu}$ whose pullback onto the worldvolume we denote by $k^{a}$. Following \cite{Blackfolds_string_theory}, we combine the asymptotic quantities of the D3-NS5 system to formulate a variational problem for the transverse fluctuations of the brane, after integrating out its intrinsic degrees of freedom. Thus, one seeks an equilibrium action - also known as equilibrium partition function in the context of relativistic fluids - that can produce the sources of the brane via a variational principle. The solutions of the equations of motion derived from this action must reproduce the correct equilibrium configurations. Such an action has already appeared in \cite{PhysRevLett.122.181601} and here we slightly generalize it to include all the sources of the D3-NS5 system as well as the running of the dilaton along the lines of \cite{2016}.

The unknown fields which can not be determined from the sole assumption of stationarity are the horizon radius $r_{0}$, the non-extremal degree of freedom $\alpha$, the field $θ$ controlling the D3 charge embedded in the five-brane and the transverse scalars. For the case studied in this paper we focus on a single transverse scalar, though the generalization is straightforward to multiple transverse scalars. The blackfold equations for the currents provide 2 global conserved charges which constrain the variations of the system. To have a well-defined variational problem we need an additional global charge. This is accomplished by specifying the statistical ensemble that we shall use and by fixing the associated thermodynamic quantity.

A peculiar property of the D3-NS5/D5 system is the non-vanishing of its temperature as a direct consequence of the limit \eqref{extremalilimit}, as can readily be seen from \eqref{thermodynamicdata} and noted in \cite{PhysRevLett.122.181601}. Since the global temperature $T$ is given by $T=|k|\mathcal{T}$\footnote{This the expected definition from the fluid point of view and can be derived by varying the action, see eq.~\eqref{Temperature}.} in the extremal limit it is a function of the transverse scalar(s) and consequently of the solution. This makes the constraint of fixed temperature unsuitable for those trajectories in the phase space with the appropriate extremal limit. Nevertheless, by looking at \eqref{thermodynamicdata} we see that the entropy density (and thus the global entropy) does vanish when we apply the limit \eqref{extremalilimit}. Therefore, we proceed to work in an ensemble for which the global entropy 
\begin{equation}\label{entropy}
     S=\int_{\mathcal{B}_{5}}\sqrt{-γ}\frac{s_{0}}{|k|}~~,
\end{equation}
is kept fixed, where we have defined the spatial part of the brane worldvolume such that $M_{6}=R\times \mathcal{B}_{5}$. That is, we let entropy flow into the system and consider the system at fixed entropy/horizon area. In this constant entropy ensemble the quantity to be extremized is the total energy $E=E(S)$ of the system, as we will show later in this section. By considering 2 systems in thermal contact in the constant entropy ensemble and following standard statistical arguments, we trivially find that one of the conditions of equilibrium is given by the equality of the respective temperatures, i.e.
\begin{equation} \label{eq:thermal_equilibrium}
    \frac{\partial E }{\partial S}|_{1}=\frac{\partial E }{\partial S}|_{2} \quad \Rightarrow \quad T_{1}=T_{2}~~,
    \end{equation}
where the subscript "1,2" label the two different systems. 

The structure of a hydrodynamic theory is determined by the global symmetries that the flow possesses. These symmetries are materialised as conservation laws. Defining $\tilde{J}_{4}=J_{4} + \star(\star j_{6}\wedge C_{2})$, we observe that the D3-NS5 system for a large\footnote{These conservation laws hold for supergravity backgrounds for which the general analysis of \cite{2016} reduces to eqs.~\eqref{conservation1}.} class of supergravity backgrounds possesses the conservation laws
\begin{equation}\label{conservationlaws}
 d\star \tilde{J}_{4}=0   \quad , \quad  d\star j_{6}=0~~.
\end{equation}
The conserved global charges arising from here, as we saw previously with purely supergravity arguments, count the number of D3 and NS5 branes enclosed by 6-dimensional and 4-dimensional surfaces, respectively. For another choice of the background or beyond the leading order analysis considered here, the above conservation laws can be modified by terms inducing a mixing between them. In the language of \cite{Gaiotto_2015}, we are facing a structure of generalized global p-form symmetries. In this context, one searches for a symmetry-based formulation of a hydrodynamic theory whose underlying microscopic degrees of freedom are bound states of D3 and D5/NS5 branes, generalizing previous works \cite{Armas_2018}. At the very least, this requires the definition of an operation - an appropriate product - between 2 higher form symmetries of respective degrees $p$ and $p+2$, corresponding to conservation laws for the higher-spin currents (in our case, these currents are $\tilde{J}_{4}$, $j_{6}$). The resulting structure is another example of a kind of higher-group symmetry \cite{https://doi.org/10.48550/arxiv.2202.04655} arising in the context of supergravity.

Let us consider the coupling of each of the p-form and (p+2)-form symmetries to background gauge fields, for the case $p=3$. For the specific supergravity setting we will consider, these gauge fields are phrased as a 6-form $\tilde{B}_{6}$ (see below) coupled to the five-branes as well as the usual RR 4-form $C_{4}$ coupled to the D3 branes. Note that we will not attempt a general treatment of the product structure outlined above. Rather, we will focus on the simplest possible case for which there is no mixing in the transformations of the gauge fields and restrict our attention to stationary configurations. The worldvolume effective action we write must reproduce the conservation laws \eqref{conservationlaws} via the requirements of background gauge invariance.

Taking into account all the above considerations, the action of the D3-NS5 black brane in equilibrium under variations that keep entropy fixed is given by
\begin{equation}\label{action}
    I=\int_{M_{6}} \sqrt{-γ}ε   -Q_{5}\int_{M_{6}}P[\Tilde{B}_{6}] - \tilde{Q}_{3}\int_{M_{4}}P[C_{4}] ~~,
\end{equation}
extremized under constant global charges $Q_{5}, \tilde{Q}_{3}$. With the symbol $P$ we denote the pullback of the background gauge potentials onto the worldvolume. For the 6-form we defined
\begin{equation}
    \tilde{B}_{6}=B_{6}-C_{2}\wedge C_{4}~~,
\end{equation}
where $B_{6}$ is introduced via
\begin{equation}\label{6form}
    dB_{6}=e^{-Φ}\star H_{3}+C_{4}\wedge F_{3}~~.
\end{equation}
In what follows, we split the worldvolume into 2 submanifolds $M_{\parallel}=M_{4}$,  $M_{\bot}$ with respect to the directions parallel and transverse to the D3 branes, denoted for later convenience as $t,x,y,z$ and $ω,φ$ respectively. Concentrating on stationary configurations, the velocity can always be aligned with a timelike Killing vector $k^a$ and be expressed as $u^{a}=\frac{k^{a}}{|k|}$ \cite{Caldarelli_2009}. In addition we assume that the 1-forms $v,w$ are given by a standard stationary ansatz which aligns each of them with the 2 transverse directions (see \eqref{goldstone}). Notice though that this is not the most general ansatz for stationary D3 branes inside the NS5. Let us in addition define $γ_{\bot}^{a b}=v^{a}v^{b}+w^{a}w^{b}$. It is then straightforward to compute the complete variation of the action \eqref{action}. The vanishing of the variation of the global entropy $δS=0$ implies
\begin{equation}\label{constraint1}
     \frac{1}{2}(γ^{a b}+u^{a}u^{b})δγ_{a b}+3\frac{δr_{0}}{r_{0}}+\frac{δ(\cosh{α})}{\cosh{α}}=0~~.
\end{equation}
Given the variation of the field $θ$ obtained from the contraint of constant global charge $\tilde{Q}_{3}$ 
\begin{equation}
    δ(\tan{θ})=-\frac{δC_{2ωφ}}{\sqrt{γ_{\bot}}}e^{-\frac{φ}{2}} -\frac{γ_{\bot}^{a b}δγ_{a b}}{2}\tan{θ} -\tan{θ}\frac{δφ}{2}~~,
\end{equation}
for the first term in \eqref{action} we find
\begin{align}\nonumber
    δ\left(\int_{M_{6}} \sqrt{-γ}ε\right)&=\int_{M_{6}}\frac{\sqrt{-γ}}{2}C\left(\frac{r_{0}^{2}}{2}γ^{a b}-r_{0}^{2}u^{a}u^{b}-r_{0}^2\sinh ^2{α}\sin ^2{θ}γ_{\bot}^{a b}+r_{0}^2\sinh ^2{α}γ^{a b}\right)δγ_{a b}\\& \label{1}
     \qquad  -\int_{M_{6}} \sqrt{-γ}\left(\frac{1}{\sqrt{γ_{\bot}}}Q_{5}Φ_{3}\right)δC_{2ωφ}+\int_{M_{6}} \sqrt{-γ}\left(\frac{C}{2}r_{0}^2\sinh^2{α}\cos^2{θ}\right)δφ~~,
\end{align}
where we defined $Φ_{3}=\sin{θ}\tanh{α}$ and for future convenience we also define $\Phi_{5}=\cos{\theta}\tanh{\alpha}$. Varying the last 2 terms of the action \eqref{action} one trivially finds
\begin{equation}\label{2}
 δ \left( -Q_{5}\int_{M_{6}}P[\Tilde{B}_{6}]\right)= - \int_{M_{6}}\sqrt{-γ}\left(\frac{Q_{5}}{\sqrt{-γ}}\right) δ(\Tilde{B}_{6tx..ωφ})~~,
\end{equation}
and
\begin{equation}\label{3}
   δ \left( -\Tilde{Q}_{3}\int_{M_{4}}P[C_{4}] \right)=- \int_{M_{6}}\sqrt{-γ}\left(\frac{\tilde{\mathcal{Q}}_3}{\sqrt{-γ_{\parallel}}}\right)δC_{4txyz}~~,
\end{equation}
where in the last equality we defined the 3-charge density such that
\begin{equation}\nonumber
    \int_{M_{\bot}}\sqrt{γ_{\bot}}\tilde{\mathcal{Q}}_3= \Tilde{Q}_{3}~~.
\end{equation}
Adding equations \eqref{1}, \eqref{2}, \eqref{3} we find that the complete variation of the action \eqref{action} is given by
\begin{equation}
   δΙ=-\int_{M_{6}} \sqrt{-γ}\left( \frac{1}{2}T^{a b}δγ_{a b}+j_{φ}δφ+J_{2}^{ω φ}\delta C_{2 ω φ}+ \tilde{J}_{4}^{t x y z}\delta C_{4 t x y z}+j_{6}^{tx..ωφ}\delta \tilde{B}_{6 tx..ωφ}\right)~~,
\end{equation}
as required. Demanding the action to be invariant under the gauge transformation 
\begin{equation}
    {δ\Tilde{B}_{6}}=dλ_{5}~~,
\end{equation}
for an arbitrary 5-form $λ_{5}$ we find upon partial integration that the current $j_{6}$ is conserved. Similarly, from the invariance of the action under the gauge transformation
\begin{equation}
    δC_{4}=dκ_{3}~~,
\end{equation}
for an arbitrary 3-form $κ_{3}$ we get as a consequence the conservation of the modified current $\tilde{J}_{4}$. Moreover, we may obtain the temperature of the black brane using the variational principle
\begin{equation}\label{Temperature}
    T_{NS5}=\frac{1}{\sqrt{-γ}}\frac{δ I}{δS}=\mathcal{T}|k|~~.
\end{equation}
As it will be relevant for later parts of this paper, we note that one may move to the canonical ensemble via a Legendre transformation
\begin{equation}\label{Legendre}
    I_{T}=I-\int_{M_{6}}\sqrt{-γ}s_{0}\mathcal{T}~~.
\end{equation}
This action should be extremized under constant global temperature $T$ and charges $\tilde{Q}_{3}$, $Q_{5}$.
 
\subsection*{Smarr relation}
We now proceed to derive the remaining global conserved charges of the stationary D3-NS5 configurations described previously. The total energy of the D3-NS5 black hole is the conserved charge associated with the differomorphism invariance of the worldvolume effective action along the timelike Killing direction. The corresponding local conserved current reads
\begin{equation}\label{conservedcurrents}
    P_{k}^{μ}=Τ^{μ ν}k_{ν}+\frac{1}{2!}J_{2}^{μ ν}C_{2 λ ν}k^{λ}+\frac{1}{4!}\tilde{J}_{4}^{μ μ_{1} μ_{2} μ_{3}}C_{4 λ μ_{1} μ_{2} μ_{3}}k^{λ}+\frac{1}{6!}j^{μ μ_{1}...μ_{5}}\tilde{B}_{6 λ μ μ_{1}...μ_{5}}k^{λ}~~.
\end{equation}
In fact, the above expression is valid for an arbitrary killing vector $k^\mu$ \cite{Armas_2018}. Equivalently, one may derive \eqref{conservedcurrents} by combining the equations of motion \eqref{blackfoldequations}, \eqref{conservation1}.  For a timelike killing vector $k^{μ}$, and defining the unit vector normal to a constant time-slice of spacetime as $η_{μ}=\frac{k_{μ}}{\sqrt{-g_{t t}}}$, the total energy is given by
\begin{equation}
    E=\int_{\mathcal{B}_{5}}dV_{5}P^{μ}_{k}η_{μ}~~,
\end{equation}
where $dV_{5}$ is the volume form along the spatial part of the worldvolume. Evaluating the formula above for the configurations of interest we obtain
\begin{equation}\label{totalenergy}
    E=\int_{\mathcal{B}_{5}}\sqrt{-γ}ε - Q_{5}\int_{\mathcal{B}_{5}}P[\Tilde{B}_{6}] - \int_{ \mathcal{B}_{5}}\Tilde{\mathcal{Q}}_{3}\sqrt{γ_{\bot}}P[C_{4}]~~.
\end{equation}
 Upon (minus) Wick-rotating and integrating the Euclidean time over a period $β=\frac{1}{T}$, we observe that extremizing the action \eqref{action} is equivalent to extremizing the total energy of the system, as previously mentioned.
 
 In order to extract the potentials conjugate to the charges $\Tilde{Q}_{3},Q_{5}$ we may again use the action \eqref{action}. The variational definition reads
\begin{equation}
    \Tilde{Φ}_{D3}=\frac{δ\mathcal{I}}{δ\Tilde{Q}_{3}}|_{S,Q_{5}}\qquad,\qquad Φ_{NS5}=\frac{δ\mathcal{I}}{δQ_{5}}|_{S,\Tilde{Q}_{3}}~~,
\end{equation}
where we defined the Wick-rotated action such that $I=(-iβ)\mathcal{I}$. Calculating the variations explicitly, we arrive at
\begin{gather}
   \Tilde{Φ}_{D3}=\frac{\int_{\mathcal{B}_{5}} Φ_{3}\sqrt{-γ}}{\int_{M_{\bot}}\sqrt{γ_{\bot}}}-\int_{\mathcal{B}_{3}}P[C_{4}]~~, \\
   Φ_{NS5}=\int_{\mathcal{B}_{5}}\sqrt{-γ}Φ_{5}-\int_{\mathcal{B}_{5}}\sqrt{-γ}Φ_{3}\frac{\int_{M_{\bot}}P[C_{2}]}{\int_{M{\bot}}\sqrt{γ_{\bot}}}-\int_{\mathcal{B}_{5}}P[\tilde{B}_{6}]~~.
\end{gather}
It is easy to see that the following identity holds
\begin{equation}
   E-\Tilde{Φ}_{D3}\Tilde{Q}_{3}-Φ_{NS5}Q_{5}=\int_{\mathcal{B}_{5}}\sqrt{-γ}\frac{3}{2}Cr_{0}^2=\frac{3}{2}TS~~.
\end{equation}
Given the above identity, we arrive at the Smarr relation
\begin{equation}
  E=\frac{3}{2}TS+\Tilde{Φ}_{D3}\Tilde{Q}_{3}+Φ_{NS5}Q_{5}~~,
\end{equation}
in agreement with the one obtained in \cite{Cohen_Maldonado_2016}.

\section{Non-extremal five-branes in mass-deformed  $AdS_{5}\times S^{5}$}\label{3.0}
In this section we discuss the classification of supersymmetric vacua of the $\mathcal{N}=1$* theory following \cite{Polchinski:2000uf}. This is followed by introducing the perturbative background of mass-deformed $AdS_{5}\times S^{5}$, which is the starting point for constructing supergravity duals to the various vacua. At the end of this section we explicitly solve the blackfold equations for non-extremal D3-NS5 branes in this background, giving evidence for the existence of black holes with $\mathbb{R}^3\times \mathbb{S}^2\times \mathbb{S}^3$ horizon topology. We also study the metastability properties of these solutions. 

\subsection{Polchinski-Strassler vacua}\label{3.1}
The field content of $\mathcal{N}=4$ with gauge group $SU(N)$ consists of 6 scalars, 4 Weyl fermions and 1 vector, all transforming in the adjoint representation of $SU(N)$. In the language of $\mathcal{N}=1$ the fields are organized in a vector multiplet V containing the vector $A_{μ}$ and the gaugino $ψ_{4}$, as well as three chiral multiplets $Φ_{i}$ of the form
\begin{equation}
    Φ_{i}=(ψ_{i},φ_{i}) \quad,\quad i=1,2,3~~,
\end{equation}
where $φ$ are 3 complex scalars. We obtain the $\mathcal{N}=1$* theory by adding to the superpotential diagonal mass terms for the chiral multiplets, which upon a rescaling of the fields can always be parametrized by one parameter, namely
\begin{equation}\label{massterms}
    δW=m_{i j} \textit{Tr}(Φ_{ι}Φ_{j}) \quad,\quad m_{i j}=mδ_{ι j}~~.
\end{equation}
The vacuum structure of this theory has been extensively studied \cite{Vafa_1994,Donagi_1996,Dorey_1999,2019}. Varying the classical superpotential yields the F-term equations for the vacua
\begin{equation}\label{FTERM}
    [φ_{ι},φ_{j}]=-mε_{i j k}φ_{k}~~ .
\end{equation}
For gauge group SU(N) the solutions of the above equations are given by N-dimensional, generically reducible, representations of SU(2). Each vacuum is therefore specified by a partition of N according to
\begin{equation}\label{partition}
    \sum_{d}d k_{d}=N~~,
\end{equation}
where the dimension of an irreducible representation is denoted by d and the non-negative integer $k_{d}$ counts the frequency of its appearance in the partition. The simplest solutions to the above equation correspond to the case for which there is no summation, such that $k_{d}=\frac{N}{D}$ for some divisor D of N. These vacua preserve an $SU(k_{d})$ gauge group classically, while quantum-mechanically they split into $k_{d}$ distinct vacua with completely broken gauge group, exhibiting a mass gap \cite{Donagi_1996}. We will refer to them as the massive vacua. Among them one distinguishes the Higgs vacuum, corresponding to the unique N-dimensional representation ($k_{d}=1$), as well as the vacuum with $k_{d}=N$ which at the quantum level splits into the N confining vacua. On the other hand, vacua obtained via a partition of N with more than one term in \eqref{partition} will always contain at least one unbroken U(1) gauge factor and are thus Coulomb vacua.

For the holographic description of the field theory vacua Polchinski and Strassler, inspired by the work of Myers on the effective action of non-abelian D-branes \cite{Dielectic-branes}, proposed a mapping to configurations of D3 branes polarized into five-branes, in an asymptotically $AdS_{5}\times S^{5}$ spacetime. The scalars transverse to the worldvolume of the brane correspond to the field theory scalars $φ_{ι}$. According to \eqref{FTERM} the D3 branes are non-commutatively expanded to form spherical shells \cite{Dielectic-branes}. The polarization process is triggered by the addition of mass deformations to $\mathcal{N}=4$ which on the gravity side amounts to turning on 3-form perturbations on the $S^{5}$. Their 7-form duals couple to the D3 branes, forcing them to expand and wrap an equator of the $S^{5}$, effectively formulating an Abelian five-brane with D3 brane charge in the static configuration $R^{1,3}\times S^{2}$. More precisely, the holographic prescription of \cite{Polchinski:2000uf} maps the vacua of the field theory to the partition of the total D3 charge N into spherical shells each carrying a D3 charge $N_{3,I}$ and $(p_{I},q_{I})$ five-brane charges, where $p_{I}$ and $q_{I}$ are the numbers of NS5 and D5 branes carried by the $I$ shell, respectively. We have
\begin{equation}\label{SUM}
    \sum_{I}N_{3,I}=N ~~.
\end{equation}
 In order for the above partition to be a viable analogue of \eqref{partition} on the gravity side one has to go beyond the probe level and move a number of order N of D3 branes, from the AdS origin all the way to an expanded shell at a different location in the throat. To this end, an important observation is the fact that the potential felt by a probe is independent of the warp factor encoding the distribution of the D3 charge in directions transverse to the worldvolume, and thus remains unaltered when the brane lives in the background sourced by one or multiple shells in the configuration $R^{1,3}\times S^{2}$. This suggests that each of the expanded shells in \eqref{SUM} minimizes its own potential and can be placed at a different non-zero AdS radius. Therefore, a  generic massive vacuum preserving classically an SU(p) gauge symmetry is described by a single fully-expanded shell made out of $p$ coincident D5 branes each carrying a D3 charge $q$, such that  $N=N_{3}=pq$ . This stems from the property of (curved) D-brane physics according to which $k$ coincident branes give rise to a low energy field theory with an enhanced SU($k$) gauge symmetry. It was furthermore argued in \cite{Polchinski:2000uf} that these vacua have an equivalent description in terms of $q$ NS5 branes each carrying D3 charge $p$\footnote{We note that there are puzzling issues associated with the precise mapping of brane configurations to field theory vacua \cite{Kinar_2001}, which however do not change the essence of our discussion here.}. On the other hand, distributions of brane sources arranged in multiple shells across the AdS radial coordinate correspond to a partition of N with multiple terms in \eqref{SUM} and are expected to be associated with the holographic description of the Coulomb vacua.

 Focusing on a single shell of NS5 or D5 kind, the effective supergravity description of the brane configuration we employ here is meaningful in the regime $N_{5},N_{3}\gg 1$, where the number of coincident five-branes is denoted by $N_{5}$ from now on, whenever the hierarchy of scales \eqref{hierarchy} can be achieved. Concentrating on the NS5 case for clarity, we search for solutions which asymptote for large values of the radial coordinate to the perturbative background of \cite{Polchinski:2000uf,Friedman} and whose near-horizon geometry and fields are given in \eqref{branegeometry} at zeroth order in the small derivative expansion. The matching of the 2 geometries in this long-wavelength regime is provided by a set of 6-dimensional differential equations \eqref{blackfoldequations}, \eqref{conservation1}. In sections \ref{3.3} and \ref{Non_extremal_PS} we will see that in the extremal limit there exist stationary solutions in agreement with the DBI results of \cite{Polchinski:2000uf}, while in the non-extremal regime these solutions exist until a critical value of the horizon area. Our matched asymptotic expansion scheme shares certain qualitative features with the interpolating metric proposed in \cite{Polchinski:2000uf}, though the crucial difference is that here the matching is by construction in agreement with the supergravity equations, including all the supergravity fields. The price to be paid is that we must start with a thin five-brane shell satisfying $N_{3}\ll N$ and perturbatively correct the solution in powers of $\frac{N_{3}}{N}$. Note that in the planar limit such a perturbative expansion is indeed possible. The foregoing discussion also implies that the Higgs vacuum corresponding to a single D5 brane ($N_{5}=1$), as well as the confining vacuum realized by a single NS5 brane, can not be directly embedded in the blackfold approach.
 
 An interesting limiting behaviour of \eqref{partition} is obtained in the regime of multiple spherical stacks of five-branes. There exists classes among those brane configurations whose generic building block is a thin enough shell and thus can be associated to the the type of solutions we are considering in this work. The significance of these solutions is reinforced by the symmetric cancellations of interactions between distinct shells emerging from \cite{Polchinski:2000uf}, suggesting that each shell can be treated independently. We note, however, that such a statement has to be checked using the explicit extremal solution obtained after taking into account the backreaction of the brane on the geometry and the fluxes, and does not hold once temperature flows into the system.
 In the limit of very large $N\rightarrow \infty$  one could view the dimension of the representation $d$ as a continuous variable $x$ and assign to it a smooth distribution $k(x)$. Equation \eqref{partition} can then be written as
 \begin{equation}
     \int_{0}^{\infty}x k(x)dx=1 ~~.
 \end{equation}
A natural realization of this limiting behaviour can be provided in the gravity side by mapping $x$ to some spacetime coordinate $ω$, $x=f(ω)$. In \cite{2019} strong evidence was presented for the appearance of a continuous distribution of five-branes with D3 charge in the near-singularity structure of the uplifted GPPZ solution. The existence of this collection of five-branes smeared across a (periodic) direction of spacetime, along with the expected screening behaviour of probe strings extended in the geometry, supports the claim of \cite{2019} that the solution in \cite{Petrini:2018pjk,2018_uplift_GPPZ} can be dual, for a given regime of parameter space, to a set of Coulomb vacua of the gauge theory.

The complexity of finding supergravity solutions that realize the mechanism envisioned by Polchinski and Strassler can be understood by looking at symmetries. The $\mathcal{N}=4$ theory possesses an $SO(6)\sim SU(4)$ R-symmetry which rotates the fields one into another. This corresponds on the string side to the isometries in the $S^{5}$. In $\mathcal{N}=1$ notation, the symmetry is already reduced in the ultraviolet to $SU(3)\times U(1)$. The introduction of 3 equal masses results in the breaking $SU(3)\times U(1) \rightarrow SO(3)$.

 In the gravity side, spoiling the R-symmetry with the mass terms \eqref{massterms} means that all supergravity fields generically obtain angular dependence and require the implementation of an at least cohomogeneity-3 ansatz, a fact that makes the construction of exact solutions directly in 10 dimensions a difficult task at the present time. This observation further motivates the adoption of blackfold methods towards the construction of explicit perturbative solutions describing the backreacted Polchinski-Strassler branes, which we will pursue below.

\subsection{Background geometry and fluxes}\label{3.2}
In this section we introduce the mass-deformed $AdS_{5}\times S^{5}$ background. The geometry sourced by any distribution of N D3 branes aligned in the $μ$ directions is given by
\begin{equation}\label{Nsources}
		     ds^{2}=Z^{-\frac{1}{2}}η_{μ ν}dx^{μ}dx^{ν} + Z^{\frac{1}{2}}dy^{m}dy^{m},\quad  μ,ν=0,1,2,3~~,
\end{equation}
where $r^{2}=y^{m}y^{m}$ and Z any harmonic function of the transverse coordinates $y^{m}$. The solution of the type IIB  equations of motion also contains the field strength
	    \begin{equation}\label{B2}
	      F_{5}=dχ_{4}+\star dχ_{4} \quad,\quad χ_{4}=\frac{1}{g_{s}Z} dt\wedge dx^{1}\wedge dx^{2}\wedge dx^{3}~~, \\ 
	    \end{equation}
while the dilaton is trivial such that $e^{Φ}=g_{s}$ (we assume a vanishing constant RR scalar for simplicity). For the choice 
\begin{equation}
    Z(r)=\frac{L^4}{r^4}\quad,\quad L^{4}=4\pi g_{s}N α'^2~~,
\end{equation}
we recover $AdS_{5}\times S^{5}$ with all the D3 branes lying at the origin $r=0$. This solution forms the seed of the perturbative supergravity background into which the expanded branes are embedded and receives corrections weighted by positive powers of $\frac{m}{r}$, where $m$ is the mass parameter and treated as a perturbation. The first order correction comes from the 3-form dual to the mass deformation of the gauge theory
\begin{equation}
    G_{3}=-\frac{\sqrt{2}}{g_{s}}L^4d\left(\frac{1}{r^4}S_{2}\right)\quad,\quad S_{2}=\frac{1}{2} T_{mnp}y^{m}dy^{n}\wedge dy^{p} ~~.
\end{equation}
Here $T_{mnp}$ is a constant, totally antisymmetric and anti-self dual tensor encoding the masses of the fermions. In standard complex coordinates parametrizing the transverse space, it reads
\begin{equation}
     T_{3}=m\left(dz^{1}\wedge d\Bar{z}^{2}\wedge d\Bar{z}^{3}+ d\Bar{z}^{1}\wedge dz^{2}\wedge d\Bar{z}^{3}+ d\Bar{z}^{1}\wedge d\Bar{z}^{2}\wedge dz^{3}\right)~~ .
\end{equation}
The backreaction of the 3-form $G_{3}$ brings $m^2$ corrections to all the fields and in \cite{Friedman} these corrections were analytically calculated. Setting $T=0$ in the formulae derived in \cite{Friedman} we obtain the expressions for the corrected geometry and field strengths to second order. This is the background that constitutes the ultraviolet boundary condition for the class of solutions we are constructing here. The mass-deformed geometry has the form
\begin{equation}\label{backgroundgeometry}
		     ds^{2}=\left(Z^{-\frac{1}{2}}(r)+h(r)\right)(η_{μ ν}dx^{μ}dx^{ν}) + Z^{\frac{1}{2}}(r)g_{m n}dx^{m}dx^{m},\quad  μ,ν=0,1,2,3 ~~,
\end{equation}	    
where $h(r)=h_{0}=\frac{7}{24}m^2L^2$ and $g_{m n}=δ_{m n}+\mathcal{O}(m^2)$. We note that the second order corrections to the transverse geometry contribute with higher order terms in the charge/mass expansion of the action \eqref{action} and for the most of this work will be neglected. This is also the case for the second order correction to the dilaton. The relevant components of the corrected 4-form potential are given by
\begin{equation}
    C_{4}=\left(\frac{r^4}{L^4g_{s}}+\frac{r^2m^2}{12g_{s}}\right)dt\wedge dx\wedge dy\wedge dz~~.
\end{equation}

Let us now comment on the validity of the effective description we employ for the D3-NS5 system in the background just outlined. The scales associated with the NS5 and D3 charge are $r_{NS5}\sim \sqrt{N_{5}}\sqrt{α'}$ and  $r_{D3}\sim \ (g_{s}N_{3})^{\frac{1}{4}}\sqrt{α'}$. A generic requirement arising from \eqref{hierarchy} comes from the 3-form $G_{3}$ which is dual to the mass perturbation of the gauge theory, and whose norm scales as
\begin{equation} \label{eq:infraredsingularity}
   |G_{3}|\sim \frac{3L}{2g_{s}}\frac{m}{r}+\mathcal{O}(\frac{m^3}{r^3})~~. 
\end{equation}\\
The 3-form possesses an infrared singularity which for the Polchinski-Strassler (PS) background is unavoidable and makes the introduction of the brane sources essential to interpret it by replacing it with an expanded five-brane. Requiring 
\begin{equation}\label{G3requirement}
      r_{NS5}\ll |G_{3}|^{-1} \qquad ,\qquad r_{D3}\ll |G_{3}|^{-1}\qquad~~,  
\end{equation}
we deduce that solutions derived from the action \eqref{action} can not be trusted as $r\rightarrow 0$. Far from this region of spacetime, the characteristic length scale of the background and, for a static embedding, of the worldvolume geometry, is the AdS radius $L$. Thus we must satisfy
\begin{equation}\label{Lrequirement}
    N_{5}\ll \sqrt{g_{s}N} \qquad,\qquad N_{3}\ll N~~.
\end{equation}
For sufficiently large $g_{s}N$ i.e. in the necessary regime for supergravity to be valid, we can always satisfy these inequalities. We moreover note that for the configurations of interest $r\sim m \frac{N_{3}}{N_{5}}α'$ and then \eqref{G3requirement} implies  $\frac{N_{3}}{N_{5}} \gg N^{\frac{1}{4}}$.

A related issue concerns the zero temperature limit of our solutions. As discussed in secction \ref{Effectiveaction} the local temperature of the brane (see Eq.~\eqref{thermodynamicdata}) does not necessarily vanish when we take the limit \eqref{extremalilimit}. However, we can make it vanish by sending $N_{3}$ to infinity and paying the cost of violating the validity regime \eqref{Lrequirement}. At the same time, for a PS massive vacuum this is just the planar limit. The point here is that we approximate the PS massive vacua by starting from an "ultra-thin" (or high-temperature) regime and iteratively moving towards a fully-expanded shell of zero temperature. As a useful analogue consider constructing, using the same perturbative scheme, a Myers-Perry black hole starting from the ultra-spinning regime and iteratively approaching the non-spinning (Schwarzschild) case \cite{Emparan:2003sy, Emparan:2009vd}.

\subsection{Ansatz for stationary solutions and extremal limit}\label{3.3}
In this section we make explicit the type of solutions of \eqref{blackfoldequations}, \eqref{conservation1} we consider and show that their extremal limit is in agreement with the DBI results of \cite{Polchinski:2000uf}.

We parametrize the 6-dimensional transverse space with metric $g_{m n}$ using standard spherical coordinates which are the radial coordinate $r$ and the 5 angles $κ , λ , ψ , ω , φ$ on the $S^{5}$. The five-branes are embedded in \eqref{backgroundgeometry} in such a way that they wrap an equator of the $S^{5}$. Working in the static gauge we set
\begin{equation}\label{embedding}
  r=R,\quad   σ^{0}=t  ,\quad  σ^{1}=x^{1} , \quad σ^{2}=x^{2},\quad σ^{3}=x^{3}, \quad σ^{4}=ω ,\quad σ^{5}=φ ,\quad  κ=λ=ψ=π/2~,
\end{equation}
and the induced metric becomes
\begin{equation}
    γ_{a b}dσ^{α}dσ^{b}=\left(\frac{R^2}{L^2}+ h_{0}\right)\left( -dt^2 + (dx^{1})^2 + (dx^{2})^2 + (dx^{3})^2 \right) + L^2(dω^2 + \sin ^2{ω}dφ^2 +\mathcal{O}(m^2)).
\end{equation}

In this section and for the rest of this paper we exploit the scaling symmetries of the equations of motion to set the asymptotic value of the dilaton equal to zero and consider the D3-NS5 and D3-D5 cases simultaneously.

The unknown degrees of freedom of the D3-NS5 black brane that we are searching for are now treated as follows. The intrinsic degrees of freedom are given by the set of worldvolume fields $u^{a},v^{a},w^{a}$. The requirement of stationarity imposes strict constraints on the form of these fields. On general grounds, we can align the velocity field with the timelike killing vector $k=k^{μ}\partial_{μ}=\partial_{t}$, such that
\begin{equation}\label{velocity}
      u=u^{a}\partial_{a}=\frac{1}{\sqrt{\frac{R^2}{L^2}+ h_{0}}}\partial_{t}~~.
\end{equation}
Analogously, we employ a stationary ansatz for the 1-forms $v,w$ so that they are aligned along the directions of the $S^{2}$, i.e.
\begin{equation}\label{goldstone}
     v^{a}\partial_{a}=\frac{1}{\sqrt{γ_{ω ω}}}\partial_{ω} \quad ,\quad w^{a}\partial_{a}=\frac{1}{\sqrt{γ_{φ φ}}}\partial_{φ}~~.
\end{equation}
Projecting equation \eqref{blackfoldequations} onto the worldvolume, we arrive at the intrinsic equations
\begin{equation}\label{intrinsic}
    \nabla_{a}T^{a b}=0~~.
\end{equation}
It is a trivial exercise to verify that equation \eqref{intrinsic} is automatically satisfied using the ansatz \eqref{velocity}, \eqref{goldstone}. On the other hand, the projection of equation \eqref{blackfoldequations} onto the direction $r$ transverse to the five-brane is given by
\begin{equation}\label{extrinsic}
      K_{a b}{\!}^{r}T^{a b}=\frac{1}{2} F^{r a_{1} a_{2}}_{3}J_{2 a_{1}a_{2}}+ \frac{e^{-Φ}}{6!}H^{r a_{1}...a_{6}}_{7}j_{6 a_{1}...a_{6}}+\frac{1}{4!}\tilde{F}^{r a_{1}..a_{4}}_{5}J_{4 a_{1}..a_{4}} + j_{Φ}\partial^{r}Φ ~,~~
 \end{equation}
where $K_{a b}{\!}^{r}$ are the components of the second fundamental tensor (extrinsic curvature) along the transverse direction $r$. The dynamical degrees of freedom, namely the horizon radius $r_{0}$, the non-extremality field $α$, the field $θ$ and the transverse scalar $R$ are now determined by the equations \eqref{extrinsic} and \eqref{conservation1}, supplemented with the constraint of constant global entropy. Notice that the conserved charges \eqref{definitionQ5}, \eqref{definitionQ3} stemming from \eqref{conservation1} combine to provide us with a global solution for $θ$ via
\begin{equation}\label{tan}
    \tan{θ}=e^{-\frac{φ}{2}}\left(\frac{\tilde{Q}_{3}}{Q_{5}}\frac{1}{4\pi L^2}+\frac{3m}{2R}L^2\right)~~.
\end{equation}

\subsection*{Extremal limit}
For clarity, we choose to work directly with the action \eqref{action}. Its extremal limit is given by
\begin{equation}\label{extremalaction}
  I=Q_{5}\int_{M_{6}}\sqrt{γ}e^{\frac{φ}{2}}\sqrt{(1+\tan ^2{θ})} - Q_{5}\int_{M_{6}}P[\Tilde{B}_{6}] - \Tilde{Q}_{3}\int_{M_{4}}P[C_{4}]~~, 
\end{equation}
where the field $θ$ is given in \eqref{tan}. This action has the same form as the DBI action, upon identifying the proportionality constants in \eqref{quantizationconditions} as follows
\begin{equation}
Q_{5}=\frac{N_{5}}{(2\pi)^5 \alpha'^{3}}\quad ,\quad \Tilde{Q}_{3} =\frac{N_{3}}{(2\pi)^3 \alpha'^2}~~.
\end{equation}
A crucial ingredient of the PS solution is the existence of a small parameter effectively proportional to the flux $G_{3}$. This small parameter enables a treatment of the system as a perturbation of the Coulomb branch of the parent $\mathcal{N}=4$ theory \eqref{Nsources}, \eqref{B2}, and is translated to the dominance of the D3 charge density. 
Thus, in order to treat the mass as a perturbation in the far-zone one has to consider the regime where the five-brane charge is a perturbation. In our setting this means that we have to consider the regime
\begin{equation}\label{ratio}
      c=\frac{\Tilde{Q}_{3}}{4\pi L^2 Q_{5}}\gg 1~~.
\end{equation}
Here $c$ represents the ratio of the effective D3 charge density to the fivebrane charge density.

In the extremal limit we should recover a supersymmetric confuguration. The expansion parameter that supersymmetry "sees" during the probe calculation is then $ε\sim \frac{1}{c}$. Inserting into \eqref{extremalaction} all the necessary data and expanding with respect to $ε$ we find
\begin{equation}\label{PSpotential}
  \frac{I}{4\pi V Q_{5}}=\frac{2\pi Q_{5}}{\Tilde{Q}_{3}}R^2\left(R-m\frac{\Tilde{Q}_{3}}{4 \pi Q_{5}}\right)^2~~,
  \end{equation}
where $V$ is the volume of the 4 Minkowski directions. In this way we recover the Polchinski-Strassler potential for NS5 shells in the configuration $\mathbb{R}^{1,3}\times \mathbb{S}^{2}$ directly in supergravity. There are 2 non-zero equilibrium configurations, one stable 
\begin{equation}\label{solution}
    R_{min}=m \frac {N_{3}}{N_{5}}\pi  α'~~,
\end{equation}
and one unstable $R_{max}=\frac{R_{min}}{2}$. One of our main purposes is to examine the metastability of the vacuum \eqref{solution}. It is straightforward to check that the extremal limit of \eqref{extrinsic}, \eqref{tan} faithfully reproduces the equations of motion derived by the DBI action in \cite{Polchinski:2000uf}.

\subsection{Moving away from extremality}\label{Non_extremal_PS}
Having established the consistency of the Polchinski-Strassler polarized branes with the blackfold equations, we now move to the non-extremal case. Using \eqref{thermodynamicdata} the action can be written
\begin{equation}\label{nonextremalaction}
     I=Q_{5}\int_{M_{6}}\sqrt{-γ}e^{\frac{φ}{2}}\sqrt{(1+\tan ^2{θ})}G(α) - Q_{5}\int_{M_{6}}P[\Tilde{B}_{6}] - \Tilde{Q}_{3}\int_{M_{4}}P[C_{4}]~~.
\end{equation}
Here, the function G of the field $α$ is equal to
\begin{equation}
    G(α)= \frac{3+2\sinh^2{α}}{2\cosh{α}\sinh{α}}~~,
\end{equation}
and encodes the information about the thermally excited worldvolume of the five-brane. In turn, the off-shell constraint of constant global entropy leads to
\begin{equation}\label{constantS}
  \sinh^8{α}+\sinh^6{α}=\frac{E(R)^6(1+\tan ^2{θ})^{3}}{\hat{S}^{4}}  \quad, \quad E(R)=\frac{R^2}{L^2}+\frac{7m^{2} L{^2}}{24}~~,
\end{equation}
where $\hat{S}$ is a conveniently normalized entropy, given by
\begin{equation}
    S=a_{2}\hat{S}=V_{3} \frac{A(N,N_{5})}{L^3}\hat{S}\quad, \quad A(N,N_{5})=\sqrt{2}N^{\frac{5}{4}}N_{5}^{\frac{3}{2}}\pi^{-\frac{7}{4}}~~,
\end{equation}
and $V_{3}$ is the volume of the spatial Minkwoski directions. The field $θ$ controlling the D3 charge transferred to the five-brane is given by \eqref{tan}. In order to analytically solve the system for $r_{0}$, $α$, and $R$, it is inevitable to recur to perturbative methods. It is desirable that our small parameter vanishes at extremality. Since the holographic duals of the field theory vacua are zero entropy limits of (collections of) the non-extremal D3-NS5 black branes, we consider a regime of small entropy, which can be thought of as a near-extremal regime. Any quantity of interest will be expressed as a series in powers of the entropy. 

Using \eqref{tan}, \eqref{constantS} we compute 
\begin{equation}\label{blackfoldinformation}
    G(α)=1+\frac{\hat{S}}{E(R)^{\frac{3}{2}}(1+\tan ^2{\theta})^{\frac{3}{4}}} -\frac{1}{8}\frac{\hat{S}^2}{E(R)^{3}(1+\tan^2{\theta})^{\frac{3}{2}}}+\mathcal{O}(\hat{S}^{3})~~.
\end{equation}
 Defining in particular the density $s=\frac{S}{V_{3}}$, the entropy corrections to the extremal polarized branes can be written as a series in powers of the dimensionless ratio $\frac{s}{m^3}$, whose coefficients are functions of $N_{3},N_{5}$ and $N$. In the solutions derived below we focus on the dominant coefficients in the regime where $c$ is large. With the result \eqref{blackfoldinformation} at our disposal and substituting all the necessary data into the equation of motion \eqref{extrinsic} derived from the action \eqref{nonextremalaction} we find the following 2 solutions 
\begin{gather}
 R_{max}=\frac{mcL^2}{2}\left(1+\frac{s}{m^3}\frac{4}{ 
 Ac^{\frac{5}{2}}}+\mathcal{O}(s^2)\right)~~,\\
 R_{min}=m c L^2\left(1-\frac{s}{m^3}\frac{1}{ 
 Ac^{\frac{5}{2}}}+\mathcal{O}(s^2)\right)~~.\label{stable}
 \end{gather}
Clearly, these correspond to the thermally-corrected PS solutions for the unstable maximum and the metastable minimum. When setting $S=0$ we recover the results of Polchinski and Strassler \cite{Polchinski:2000uf} given already in \eqref{solution} and below. Both solutions have horizon topology $\mathbb R^{3}\times \mathbb S^{2} \times \mathbb S^{3}$.  We observe that the dominant effect is that the 2 extrema approach each other. The existence of the stable solutions \eqref{stable} which are continuously connected with the PS polarized branes constitutes strong evidence in favour of the claim that the PS vacua are not gapless, but persist to exist as temperature flows into the system and are thus metastable. In fact, there exists a critical value of the entropy until which the metastable state survives. We can give an estimation of the critical entropy by locating the value of the transverse scalar for which the 2 extrema collide. We find
 \begin{equation}\label{criticalentropy}
      \hat{S}^{*}\approx \eta (m L)^3 c^\frac{5}{2}\quad,\quad \eta=0.0962~~.
 \end{equation}

\begin{figure}
\centering
\begin{subfigure}{.5\textwidth}
  \centering
  \includegraphics[width=.9\linewidth]{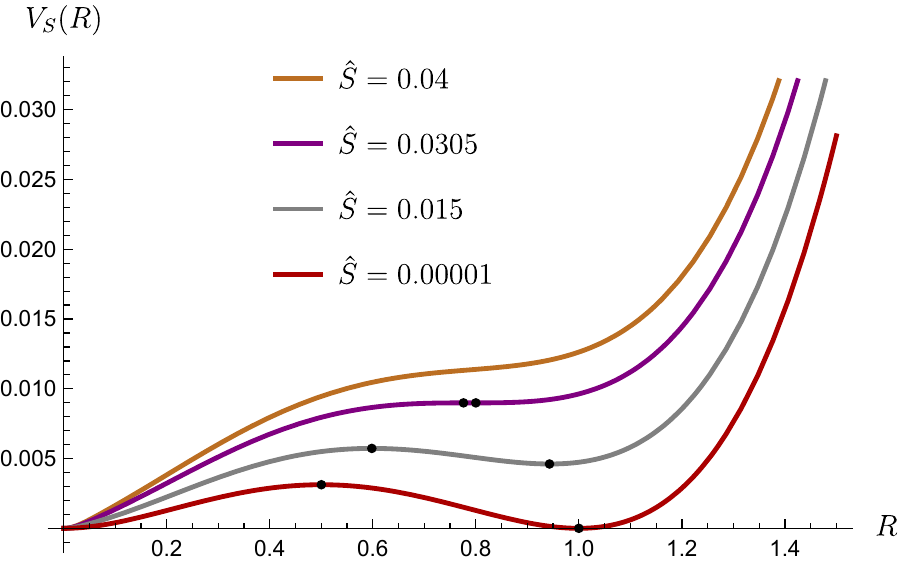}
  \caption{}
  \label{effectivepotential}
\end{subfigure}%
\begin{subfigure}{.5\textwidth}
  \centering
  \includegraphics[width=.9\linewidth]{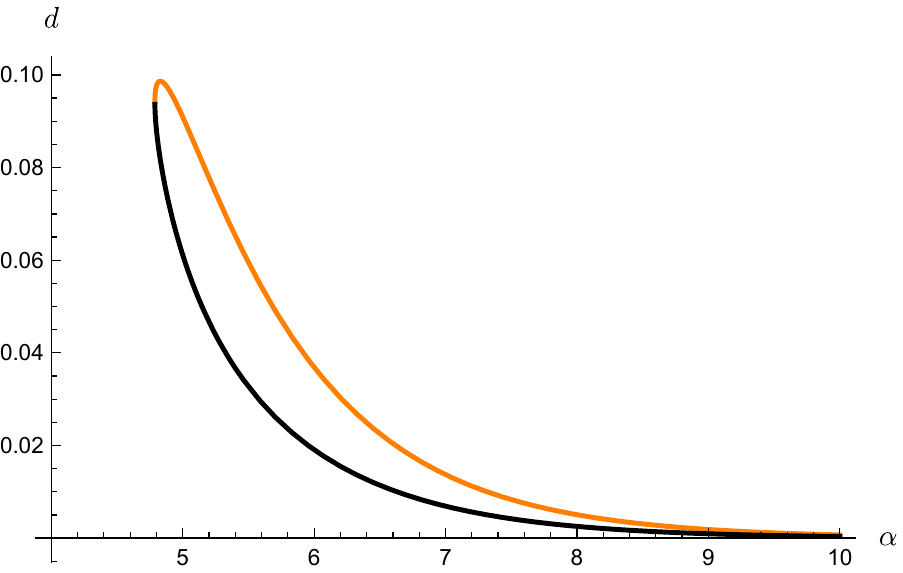}
  \caption{}
  \label{ratioD}
\end{subfigure}
\caption{\textit{a) Effective potential for a choice of parameters such that $c=10$, where for illustrative reasons we pick m so that the non-zero extremal minimum is located at one in $L=1$ units. The black dots represent the non-trivial extrema. The purple curve captures the potential felt by the brane for a value of the entropy close to the critical one. b) Plot of the ratio $d$ comparing the scale associated to the Schwarzschild radius with the radius $R$ set by the equations of motion of the brane, as a function of the non-extremality parameter $α$, for the choice $c=30$ and keeping the same convention for the mass perturbation as in \ref{effectivepotential}}.}
\label{fig:test}
\end{figure}

 We may use the expression for the function $G(α)$ \eqref{blackfoldinformation} to determine the entropy corrections because of the thermalization of the D3-NS5 system in the PS background directly in the effective potential $V_{S}=\frac{I}{4\pi Q_{5} V_{4}}$.  Observe the main modification that the presence of a non-zero brane horizon brings to the probe potential. At non-zero $S$ the leading terms in the charge expansion do not cancel, giving rise to the dominant entropy correction \eqref{PSpotential}
\begin{equation}\label{correctionPS}
   ΔV=\hat{S}\frac{R L}{\sqrt{c}}~~.
\end{equation}
An analogous effect due to the presence of a non-zero background horizon was also observed in \cite{Friedman}. For concreteness, in figure \ref{effectivepotential} we show the plot of the effective potential keeping only the dominant correction \eqref{correctionPS}.

 Conclusively, in all cases we considered in the regime where the D3 charge dominates we encountered the same pattern: as we turn the entropy on the metastable vacuum continues to exist, gets lifted and moves towards the unstable maximum. Further raising the entropy results in a collision of the 2 extrema at a critical value of the entropy. The dominant term in the expansion of this quantity is given in \eqref{criticalentropy}. Right after the merger, the metastable state is lost. The picture emerging from this analysis is qualitatively similar with that of non-extremal antibranes at the tip of the Klebanov-Strassler throat in ref. \cite{PhysRevLett.122.181601}, though here there are certain differences which we comment below.
 
 A clear difference concerns the mechanism responsible for the loss of the metastable state. To investigate this issue we introduce a quantity which can give us information regarding the physics near the transition point. A measure of the thickness of the black D3-NS5 state can be defined as
 \begin{equation}\label{ratiod}
   d=2l_{s}\frac{\sqrt{n\frac{\mathcal{C}}{Q_{5}}}r_{0}}{R}~~,
\end{equation}
where $n=\frac{N_{3}}{N_{5}}$, comparing the scale of the Schwarzschild radius of the black hole with the conformal radius $R$ of the NS5 shell. In figure \ref{ratioD} we depict this ratio using the equations of motion as a function of the non-extremality parameter $α$.

The black line captures the metastable branch while the orange one corresponds to the evolution of the unstable vacuum. We observe that as we thermalize the system the Schwarzschild radius of both states grows, while throughout the process the values of $d$ are always bounded from above. In the terminology of \cite{m2-m52019} we may interpret the transition point as a thin-thin merger between 2 black holes. This is in contrast with the results of \cite{PhysRevLett.122.181601} where the metastable state was lost via a thin-fat merger driven by the properties of the horizon. 

We end this discussion by noting that the analysis presented here allows us to ascertain the existence of a black hole in mass deformed $AdS_{5}\times S^{5}$ with spatial horizon topology $\mathbb R^{3}\times \mathbb S^{2} \times \mathbb S^{3}$. These configurations are continuously connected to the PS vacua by sending $S\to0$ and are thus the first example of this type of solutions directly constructed in 10-dimensional supergravity. However, one should note that these solutions are perturbative, and higher-order corrections would need to be included in order to describe exactly a class of PS massive vacua \cite{Polchinski:2000uf}. We also note that our solution is able to hide potential infrared singularities (due to brane sources, see \eqref{eq:infraredsingularity}) behind the horizon when $ S\ne0$. We will describe further features of these black holes in the next chapter. 

\subsection*{Energy and first law}
The black hole solution we are considering is characterized by the parameters $m, L, N_{3}, N_{5}$. For fixed L (or, at leading order in $\frac{N_{3}}{N}$, for fixed N) one expects on-shell a first law of the form 
\begin{equation}
    \delta E=T\delta S + \lambda \delta m + \mu _{3} \delta N_{3} +  \mu_{5} \delta N_{5}~~,    
\end{equation}
where $\lambda$ is the conjugate variable to the mass perturbation $m$ and $\mu_{3}$, $\mu_{5}$ are the chemical potentials corresponding to $N_{3}$, $N_{5}$. For fixed $N_{3}$, $N_{5}$ we recover the first law already written in \cite{Bena_2019, Dias:2019wof}. It should be noted that one expects the total five-brane charge of the soution to be zero, so that the conserved charge density $Q_{5}$ and the associated conserved number $N_{5}$ actually correspond to a dipole charge. At the same time, the total D3 charge of the solution at leading order is $N$. Once the fields are corrected in the far-zone, it is likely that this quantity receives $\frac{N_{3}}{N}$ corrections. 

As noted in section \ref{Effectiveaction}, the temperature of the five-brane can be derived by varying the action \eqref{action}. Evaluating Eq.~\eqref{Temperature} on-shell we can write the entropy expansion of the black hole temperature
\begin{equation}\label{PStemperature}
    T(S)=T_{ex}-g_{1}S+\mathcal{O}(S^2)~~, 
\end{equation}
where the extremal temperature is
\begin{equation}
    T_{ex}=a_{1}\frac{\sqrt{E(R_{min})}}{(1+\tan^{2}{\theta})^\frac{1}{4}} \quad, \quad a_{1}=\frac{1}{2\pi}\sqrt{\frac{\mathcal{C}}{Q_{5}}}~~,
\end{equation}
and the dominant contribution to the linear coefficient is $g_{1}=\frac{a_{1}}{a_{2}(m L)^2c^2}$. In turn, evaluating the (Wick-rotated) action, and thus the total energy, on the metastable solution we obtain
\begin{equation}\label{EnergyPS}
    E=T_{ex}S-\frac{g_{1}}{2}S^2 +\mathcal{O}(S^3)~~. 
\end{equation}
For zero entropy the energy is zero, as required for an extremal supersymmetric configuration.

The thermodynamic properties and relations of the black hole solution considered here hold order-by-order in the entropy expansion. As a useful check, we observe that using the energy at order $\mathcal{O}\left(S^2\right)$ the variation $\frac{\delta E}{\delta S}$ reproduces the linear entropy expansion of the temperature $T$ given by \eqref{PStemperature}, as expected. Furthermore, it is straightforward to verify that the solution found here satisfies the (on-shell) Smarr relation 
\begin{equation}
    TS=\frac{4}{3}E-\frac{1}{3}\lambda m~~,
\end{equation}
in agreement with the Smarr relation previously derived in \cite{Bena_2019, Dias:2019wof} for black holes in mass-deformed AdS.

\section{Five-branes in the high temperature phase}\label{4.0}
In this section we provide a similar analysis as in the previous section but in the high-temperature phase of the $\mathcal{N}=1$* in which the background is that of a black brane in asymptotic mass deformed $AdS_{5}\times S^{5}$ obtained perturbatively in \cite{Friedman}. The analysis will reveal the existence of black holes with spatial horizon topology $\mathbb R^{3}\times \mathbb S^{2} \times \mathbb S^{3}$ in this high temperature phase. At the end of this section we will combine the results of the previous section and that of \cite{Bena_2019} in order to construct phase diagrams.

\subsection{$AdS_{5}\times S^{5}$ black hole background}\label{4.1}
Adding temperature to brane configurations can be performed from the perspective of holography in various ways \cite{m2-m52019}. The most common approach is to thermalize the background by considering the system in the vicinity of a black hole solution. A second method focusing more directly on the properties of the brane is by heating-up the state itself and, from this perspective, this is the route we followed in the previous chapter. In this section we follow a third, more general approach to describe thermal effects in metastable states in string theory by thermalizing at the same time all the sectors of the theory. This approach has been applied in \cite{2012b} to the study of Wilson loops in AdS/CFT.

The high temperature deconfined vacuum of the $\mathcal{N}=1$* theory is dual to a black hole in the infrared, in an asymptotically $AdS_{5}\times S^{5}$ spacetime. In \cite{Friedman} the backreaction of the 3-form on the metric, five-form and the dilaton for the finite temperature case were analytically calculated. Using numerical methods the authors of \cite{Bena_2019} managed to construct a black hole solution of type IIB supergravity dual to the high temperature deconfined vacuum of the theory with horizon topology $\mathbb{R}^{3}\times \mathbb{S}^{5}$, at all orders in the mass perturbation. Their result for the effect of the fermion masses in the entropy at strong coupling is in agreement with the perturbative results of \cite{Friedman}. Employing the effective theory outlined in section \ref{2.0}, here we study analytically the D3-NS5 system on top of the high temperature deconfined vacuum by investigating stationary solutions of \eqref{blackfoldequations}, \eqref{conservation1}.

The background geometry into which we embed the five-brane is the mass-perturbed thermal D3 brane solution obtained in \cite{Friedman}, which has the expression
\begin{equation}\label{MFbackground}
    ds^2=E(r)\left (f(r) dt^2+dx^2+dy^2+dz^2\right)+\frac{L^2}{r^2}\left(\frac{dr^2}{f(r)}+r^2d\Omega^2_{5} + \mathcal{O}(m^2)\right)~~,
\end{equation}
where
\begin{equation}
f(r)=1-\left(\frac{r_{H}}{r}\right)^{4}  \quad , \quad  E(r)=\frac{r^2}{L^2}+\frac{7m^2L^2}{24}~~.
\end{equation}
The Hawking temperature of the black hole reads 
\begin{equation}
   T_{AdS}=\frac{r_{H}}{\pi L^2}+\mathcal{O}(m^2)~~.
\end{equation}
We note that we will make use of an expansion in powers of $\frac{r_{H}}{r}$ to simplify complicated expressions in the bulk. Consequently our results are valid in the region $r\gg r_{H}$, far away from the horizon. In all calculations to follow we approximate the fluxes $G_{3}$, $H_{7}$ and $F_{5}$ by their expressions close to the boundary, given in section \ref{3.2}.

Before moving on, we have to consider the requirement of separation of scales in the presence of the new scale associated with the temperature of the black hole, $r_{H}=T\pi L^2 $. In addition to the requirements in section \ref{3.2}, far away from the region $r \rightarrow 0$ with large curvature, we must satisfy
\begin{equation}
    r_{D3}, r_{NS5} \ll T_{AdS} L^2~~,
\end{equation}
which in $L=1$ units leads to
\begin{equation}\label{temperaturecondition}
    T_{AdS} \gg \left(\frac{N_{3}}{N}\right)^{\frac{1}{4}}\quad, \quad T_{AdS} \gg \frac{\sqrt{N_{5}}}{(g_{s}N)^{\frac{1}{4}}} ~~.
\end{equation}
Given \eqref{Lrequirement} these inequalities impose rather weak lower bounds on the temperatures we can examine. These bounds only say that we can not have a valid calculation for an arbitrarily small temperature of the background, since then the horizon shrinks to $r=0$. Therefore, we will not be able to approach the $T_{AdS}=0$ case in the previous section in such a way that the solution remains within the regime of validity of the method we employ.

\subsection{Extremal limit}\label{4.2}
We continue to focus on the wrapped D3-NS5 brane in the configuration $\mathbb R^{1,3}\times \mathbb S^{2}$. For the static embedding defined in \eqref{embedding} the induced metric becomes
\begin{equation}
      γ_{a b}dσ^{α}dσ^{b}=E(R)\left( -f(R)dt^2 + dx^2 + dy^2 +dz^2 \right) + L^2(dω^2 + \sin ^2{ω}dφ^2 +\mathcal{O}(m^2))~~.
\end{equation}
Our ansatz for the worldvolume fields $u^{a}, v^{a}, w^{a}$ has the same form with the one in section \ref{3.3}. Specifically, the velocity field is
\begin{equation}\label{velocityMF}
      u=u^{a}\partial_{a}=\frac{1}{\sqrt{f(R)E(R)}}\partial_{t}~~.
\end{equation}
The 1-forms $v,w$ are still given by \eqref{goldstone}. Notice that, as every other quantity of interest, these expressions receive corrections coming from the mass $\frac{m}{r}$ and temperature $\frac{r_{H}}{r}$ expansions.

Inserting the necessary data into the extremal action \eqref{extremalaction} and truncating the 2 parallel expansions at a convenient order, we arrive at the effective potential
\begin{equation}\label{MFsimplified}
    \frac{I}{4\pi V Q_{5}}=\frac{\Tilde{Q}_{3}}{4 \pi Q_{5}}\frac{R^4}{L^4}\left(\sqrt{1-\frac{r_{H}^4}{R^4}}-1\right)+\frac{2\pi Q_{5}}{\Tilde{Q}_{3}}R^2\left(R-m\frac{\Tilde{Q}_{3}}{4\pi Q_{5}}\right)^2~~,
\end{equation}
which is the same function (upon an irrelevant normalization) of the transverse scalar as the one found using the DBI action \cite{Friedman}. Furthermore, it is straightforward to check that the extrinsic equation \eqref{extrinsic} coincides with the equation of motion derived from \eqref{MFsimplified}.

 The extremal case is characterized by the existence of a maximum background temperature $T^{*}$ beyond which the metastable vacua is lost. In the next subsection we will see that such a critical temperature is also present in the non-extremal regime. Interestingly, this critical value of the temperature scales as $\sim c^{\frac{3}{4}}$ where $c$ is the large parameter expressing the dominance of the D3 charge. On the other hand, it was demonstrated in \cite{Friedman} that the horizon shrinks as a function of the mass perturbation, suggesting the existence of a critical temperature $T_{HP} \sim m$. The latter separates the theory into high and low temperature phases. By direct analogy with $\mathcal{N}=4$ one expects that here the mass parameter $m$ plays the role of the compactification radius $L$, leading to a Hawking-Page transition at the temperature $T_{HP}$. This means that the metastable five-branes can survive way above $T_{HP}$, and the corresponding supergravity solutions, if they turn out to exist in equilibrium (see \ref{4.4}), may dominate the ensemble at intermediate phases.

\subsection{Non-extremal regime}\label{4.3}
\begin{figure}
\centering
\begin{subfigure}{.5\textwidth}
  \centering
  \includegraphics[width=.9\linewidth]{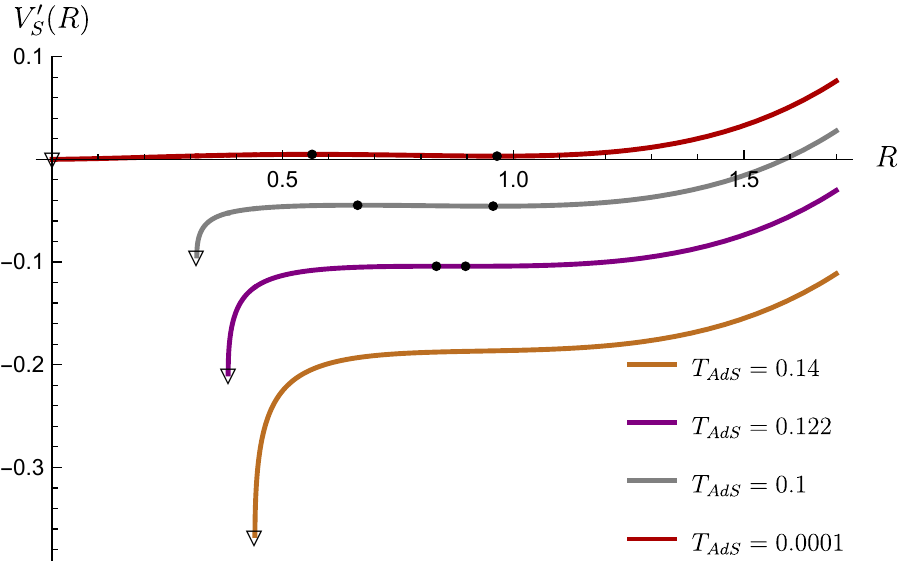}
  \caption{}
  \label{MFconstantS}
\end{subfigure}%
\begin{subfigure}{.5\textwidth}
  \centering
  \includegraphics[width=.9\linewidth]{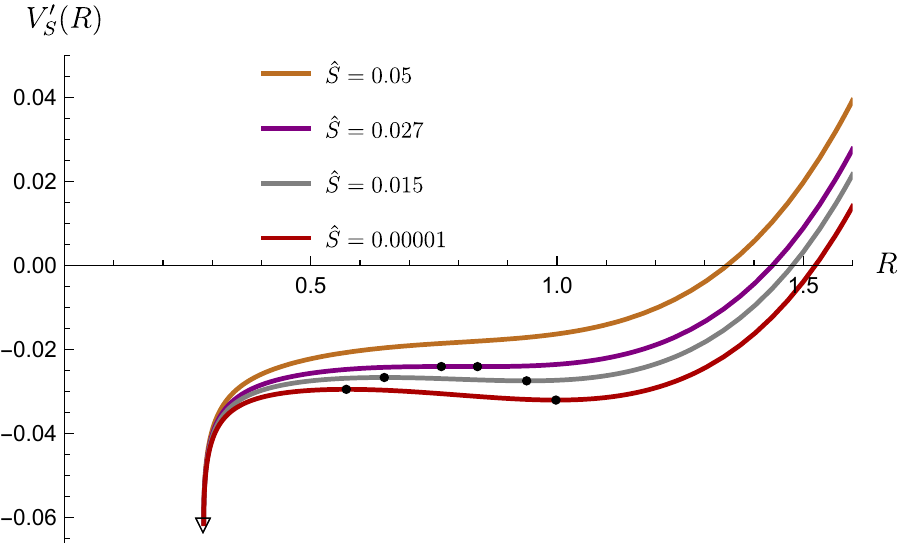}
  \caption{}
  \label{MFconstantT}
\end{subfigure}
\caption{\textit{Effective potential using the same choice of parameters and conventions as in figure \ref{effectivepotential}, for fixed entropy $\hat{S}=0.01$ (left panel) and for fixed temperature $T_{AdS}=0.09$. The purple curve captures the potential felt by the brane close to criticality. The triangle $\triangledown{}$ keeps track of the global minimum at the horizon of the background black hole. }}
\label{MFplots}
\end{figure}

 Analogously to section \ref{Non_extremal_PS} we now use the action \eqref{action} to explore non-extremal effects in the metastable polarized branes in the background of the black hole solution \eqref{MFbackground}. The effective potential of the black hole reads 
\begin{equation}\label{nonextremalMF}
     V'_{S}=c\frac{R^4}{L^2}\left(G(\alpha)\sqrt{1-\frac{r_{H}^4}{R^4}}-1\right)+\frac{R^4}{2 c L^2}G(\alpha)-mR^3 + \frac{c m^2 L^2}{12}R^2\left(7G(α)-1\right)~~.
\end{equation}
\\
Somewhat surprisingly, the leading order solution for the function $G(\alpha)$ is not affected by the background temperature. In fact, it was calculated in \eqref{blackfoldinformation}. This occurs because the constraint of global entropy \eqref{constantS} remains unaltered for the static embedding \eqref{embedding}. Substituting the solution for $G(\alpha)$ into \eqref{nonextremalMF} we have at our disposal all the entropy corrections to the effective potential \eqref{nonextremalMF}. In figure \ref{MFplots} we depict the effective potential as a function of the transverse scalar $R$ keeping only the dominant entropy corrections in two 2 different regimes which we now discuss.

Perhaps the most striking new feature arising from the non-extremal analysis of the five-brane is the simultaneous presence of 2 critical values for thermodynamic quantities associated with the background black hole and the five-brane itself. There exists no stable five-brane configuration for background temperatures greater than $T^{*}_{NE}$ nor for brane entropy greater than $S^{*}_{NE}$. More precisely, on the one hand, for any fixed horizon area of the brane which permits the formation of a metastable state, we encounter a maximum value of the background temperature until which the metastable state survives. This critical temperature  becomes a function of the entropy, as well as of $m$ and $c$. The leading extremal result $T^{*}\approx 0.7336 \pi^{-1} m c^{\frac{3}{4}}$ receives contributions coming from the presence of the brane horizon, which result in lowering the critical background temperature. On the other hand, for any fixed temperature allowing for the formation of a metastable state our findings are very close with those of the analysis in section \ref{Non_extremal_PS}, i.e. increasing the brane entropy results in a collision of the extrema after which the metastable is lost. The critical value of the entropy at which this merger occurs $S^{*}_{NE}$ has the leading behaviour of \eqref{criticalentropy} and receives contributions at order $T_{AdS}^{8}$ for large $c$.

To sum up, a general lesson learned is that the metastable D3-NS5 state ceases to exist via a merger with the unstable configuration. The merging of these 2 black holes can be provoked by a growing temperature of the (background) $AdS_{5}\times S^{5}$ black hole and/or by a slowly expanding horizon area of the branes.

We now proceed and compute thermal contributions to the on-shell value of the potential (total energy) of the metastable five-brane. These are associated to the temperature of the background black hole and the entropy of the brane. The leading thermal corrections are given by
\begin{equation}
  \frac{E_{FM}(R_{min})}{4\pi Q_{5} V_{3}}=m \hat{S} L^3 \sqrt{c} -\frac{c\pi^4}{2}T_{AdS}^4 L^6+ \mathcal{O}(ST_{AdS}^{4})~~.
\end{equation}
 We see that in the regime \\
\begin{equation}\label{BFdominance}
    \hat{S}T_{AdS}>\frac{L^3c^{\frac{1}{2}}}{2m}\pi^4 T_{AdS}^{5}~~,
\end{equation}\\
the blackfold correction prevails. Recall the condition \eqref{temperaturecondition} related to the temperature of the geometry \eqref{MFbackground}. We conclude that for sufficiently (though not arbitrarily) small temperatures $T_{AdS}$ the effects associated with the thermally excited worldvolume of the brane become important and eventually dominate the total energy of the configuration. We may observe the dominant role of the blackfold correction by checking the condition \eqref{BFdominance} for a five-brane with approximately half the critical horizon size, i.e. when $\hat{S}\approx \hat{S}^{*}/2$. For such a five-brane the internal processes of the bound state prevail for
\begin{equation}
    T_{AdS}<(0.1) D m\sqrt{c}~~,
\end{equation}
where $D$ is a $\mathcal{O}(1)$ number. This shows that the blackfold correction dominates the total energy of generic configurations even for temperatures way above the expected Hawking-Page transition.
\\
\\
So far we have concentrated our attention on the properties of the D3-NS5 bound state. However, the supergravity solution studied in this chapter consists of 2 black holes with disconnected horizons. We discuss further properties of this 2-black hole system in the next section where we construct phase diagrams in 2 distinct ensembles.

\subsection{Features of phase space}\label{4.4}
Our results indicate the existence of a new class of (multi-)black hole spacetimes in the supergravity dual of $\mathcal{N}=1$* theory, which have the interpretation of being thermal states of the dual gauge theory. Here we take a step towards a better understanding of the phase space of the theory by considering phase diagrams which take into account the perturbative supergravity solution we dealt with in this paper as well as analytic \cite{Friedman} and numerical \cite{Bena_2019} solutions found in earlier studies. We will be studying the phase space in the two ensembles that have been used in the string dual of $\mathcal{N}=1$*, namely the constant entropy $S$ and the canonical ensembles.

Prior to drawing the phase diagrams, it is necessary to study further the properties of the solution consisting of a black five-brane on top of the mass-perturbed thermal D3-brane background, which we constructed in section \ref{4.3}. This solution contains two (disconnected) horizons, each of which have their own associated temperature and are thus generically out-of-equilibrium. In order to study the phase diagram one needs to understand whether, and under which conditions, can the two horizons be in thermal equilibrium.\footnote{In the ensembles we consider, thermal equilibrium requires that the temperatures of the two horizons are equal, see Eq.~\eqref{eq:thermal_equilibrium}.} To answer this question we look at the temperature of the five-brane \eqref{Temperature} which is a natural function of the background black hole temperature $T_{AdS}$. The leading behaviour of \eqref{Temperature} is
\begin{equation}
    T^2=\frac{a_{1}^2}{c}(\tanh{\alpha})E(R)f\left(\frac{T_{AdS}}{R}\right)+\mathcal{O}(T_{AdS}^{8})~~.
\end{equation}
Focusing on the extremal limit and imposing thermal equilibrium $T=T_{AdS}$, the equation above implies
\begin{equation}
    \frac{r_{H}^{2}}{R_{min}^2}=\frac{N}{N_{3}}\left(1-\frac{r_{H}^{4} }{R_{min}^4}+...\right)~~.
\end{equation}
This resulting condition is inconsistent with the approximations we have made and the validity regime outlined in section \ref{3.2}. Indeed, we have placed the brane far from the background horizon relying on an expansion in powers of small $\frac{r_{H}}{R}$ in order to write the background temperature corrections to the probe potential. At the same time, the brane has to be thin enough in order for the leading blackfold approximation to be valid, leading to $N \gg N_{3}$. Incorporating leading finite $S$ corrections does not change this conclusion. We infer that the two black holes are out of equilibrium at ideal order in this perturbative scheme, since the thin NS5 is too hot to begin with for its temperature to be treated as a small perturbation.

It is possible that the absence of a valid regime where the two black holes are in thermal equilibrium is an artifact of the various approximations we made. For instance, one could allow for the brane to move closer to the background horizon, work with the full bulk expressions given in \cite{Friedman} and search numerically for metastable states. This scenario is however unlikely to be successful because, besides requiring accounting for additional interaction terms between the two black holes which are hard to predict without solving directly the full set of Einstein equations, we also expect it to be difficult to balance the growing gravitational force between the two horizons. Another potential solution to this issue would be to consider a five-brane of $N_{3}\sim N$ D3 charge while still being far away from the horizon $r_{H}$. However, this requires including higher order corrections in $\frac{N_{3}}{N}$ and is outside the scope of the present work. Below we study the phase diagram for solutions in thermal equilibrium and thus do not consider the construction of section \ref{4.3}.

\subsection*{Constant entropy ensemble}
Here we discuss properties of the phase space in the ensemble where the global entropy is kept fixed. The preferred configuration in the ensemble is the one with lowest energy, which is given as a function of the entropy. It is useful to define the dimensionless entropy density as follows
\begin{equation}
    \hat{s}_{m}=\frac{s_{N}}{m^{3}}\quad,\quad s_{N}=\frac{s}{N^{2}}~~,
\end{equation}
while for a dimensionless energy density we define
\begin{equation}
    \hat{\rho}_{s}=\frac{\rho}{N^2m^4} \quad, \quad \rho=\frac{E}{V_{3}}~~.
\end{equation}

\begin{figure}
\centering
\begin{subfigure}{.5\textwidth}
  \centering
  \includegraphics[width=.75\linewidth]{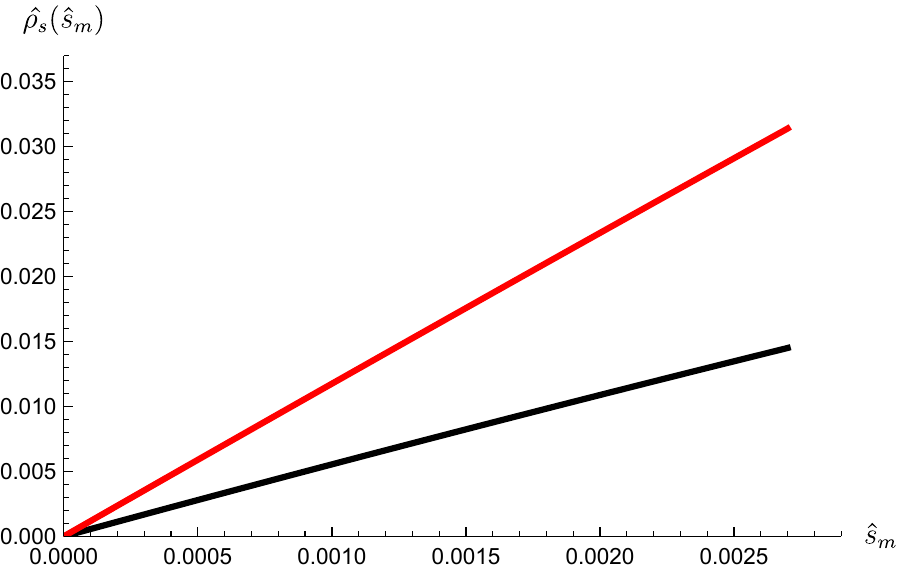}
  \caption{}
  \label{ConstantS_BF}
\end{subfigure}%
\begin{subfigure}{.5\textwidth}
  \centering
  \includegraphics[width=.75\linewidth]{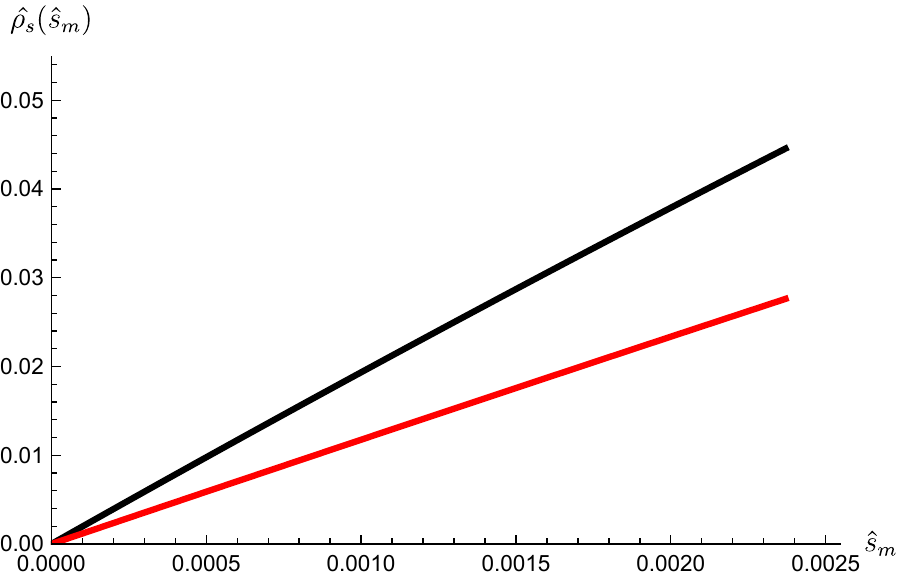}
  \caption{}
  \label{ConstantS_NUM}
\end{subfigure}
\caption{\textit{Dimensionless energy density $\hat{\rho}_{s}$ as a function of $\hat{s}_{m}$. In both plots the black line corresponds to the blackfold solution of section \ref{3.0}, while the red to the (red) branch of the numerical solution of \cite{Bena_2019} which has a zero entropy limit. We plot the 2 solutions in the entropy regime that they co-exist. In (a) the blackfold energy density is depicted for the choice $\frac{\sqrt{N_{3}}}{N_{5}}=20$ and $\frac{N_{3}}{N}=0.1$ and is the preferable configuration in the constant entropy ensemble. In (b) the choice of parameters is such that $\frac{\sqrt{N_{3}}}{N_{5}}=70$ and $\frac{N_{3}}{N}=0.05$ and the numerical solution dominates.
}}
\end{figure}
Given these definitions, the dimensionless energy density corresponding to the black D3-NS5 brane in \eqref{EnergyPS} has the leading behaviour

\begin{equation}\label{4.20}
    \hat{\rho}_{s,bf}=\frac{\sqrt{N_{3}}}{N_{5}}\frac{1}{2\sqrt{\pi}}\hat{s}_{m}-\left(\frac{N}{N_{3}}\right)^2\hat{s}_{m}^2+ \mathcal{O}(\hat{s}_{m}^{3})\quad,\quad \hat{s}_{m}<\frac{N_{3}^\frac{5}{2}}{N^2 N_{5}}\frac{\eta}{4\sqrt{\pi}} ~~.
\end{equation}
The next step is to consider the exact numerical solution found in \cite{Bena_2019}. One may move from the canonical to the constant S ensemble using \eqref{Legendre} and inverting the relation $S=S(T) \rightarrow T=T(S)$\footnote{To obtain such a relation for each of the branches of the numerical solution, we approximate the result of \cite{Bena_2019} with a polynomial for the quantity $\frac{s_{N}}{T^3}$ as a function of $\frac{m}{T}$, and numerically invert the result with respect to $T$.}. The numerical solution possesses 2 branches of solutions, one of which has a zero entropy limit. Since the blackfold solution has by definition a supersymmetric limit, we are primarily interested in the comparison of its energy density with the one obtained for the (red) branch of the numerical solution with such a limit. Recall that the phase space of the theory for fixed $N$ is described by the parameters $m,N_{3},N_{5}$. In figs. \ref{ConstantS_BF} and \ref{ConstantS_NUM} we plot the (dimensionless) energy density in this low entropy regime for two different choices of $(N_{3},N_{5})$.

We see that one can find regimes of parameter space for which the energy density of the five-brane is larger, equal or even smaller than the one of the mass-deformed thermal D3 brane. This suggests that the configuration of lowest energy depends on the particular values of $N_{3}$ and $N_{5}$. For large enough values of the entropy ($\hat{s}_{m}\geqslant \hat{s}^{1}_{m}\approx 0.03$) one must take into account the other (blue) branch of the numerical solution. From \eqref{4.20} we deduce that the blackfold solution can indeed exist in this high entropy regime. For any fixed ratio $\frac{N_{3}}{N}$ this can happen by increasing its extremal temperature (or, equivalently, the ratio $\frac{\sqrt{N_{3}}}{N_{5}}$). As a result, the energy density of the blackfold significantly increases so that the numerical solution becomes the preferable configuration. As far as the unstable branch of blackfold solutions is concerned, these have generically higher values of energy density than those of the metastable branch.

It is interesting to note that in \cite{Bena_2019} evidence was found for the supersymmetric point obtained in the limit $s\rightarrow 0$ (or $\rho \rightarrow 0$) representing the supersymmetric GPPZ solution, which has been argued in \cite{2019} to contain a smeared distribution of five-branes. The latter are extremal and as such one expects a non-zero temperature for a collection of them. Given the features of the configurations examined in this paper, with $T_{ext}$ even way above $T_{HP}\sim m$, a high temperature of this order for the GPPZ solution would not be a surprise.

\subsection*{Canonical Ensemble}
We now consider the phase diagram in the canonical ensemble (for fixed temperature). It is again useful to define dimensionless quantities.\footnote{Throughout this subsection we follow the conventions of \cite{Bena_2019}, but include the $1/N^2$ normalization in the definition of $\hat{f}$.} The dimensionless free energy density takes the form
\begin{equation}
    \hat{f}=\frac{f}{N^2 T^{4}}=\frac{\hat{\rho}-\hat{s}}{N^2}~~,
\end{equation}
where
\begin{equation}
  \hat{\rho}=\frac{\rho}{T^{4}}\quad,\quad \hat{s}=\frac{s}{T^{3}}~~. 
\end{equation}

\begin{figure}
\centering
\begin{subfigure}{.32\textwidth}
  \centering
  \includegraphics[width=0.95\linewidth]{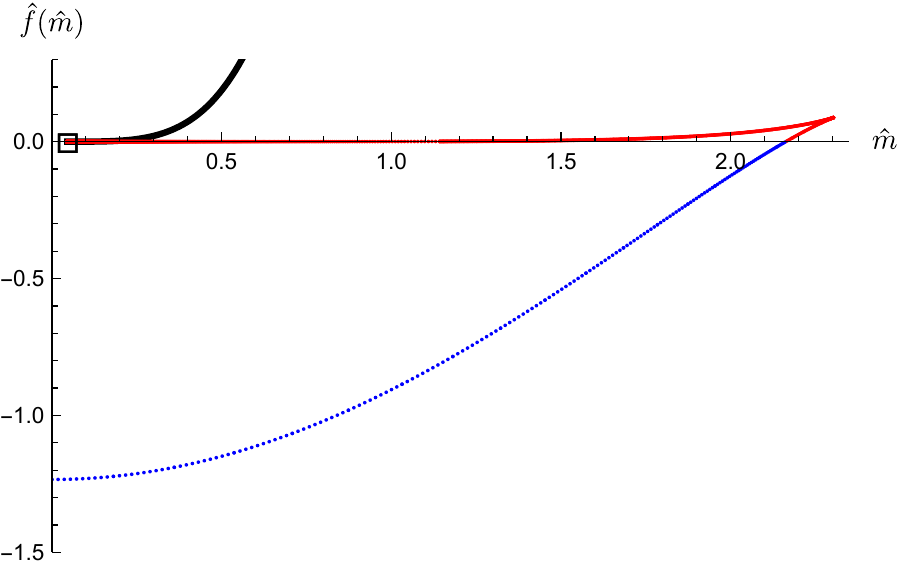}
  \caption{}
  \label{canonicalPS}
\end{subfigure}%
\begin{subfigure}{.32\textwidth}
  \centering
  \includegraphics[width=0.95\linewidth]{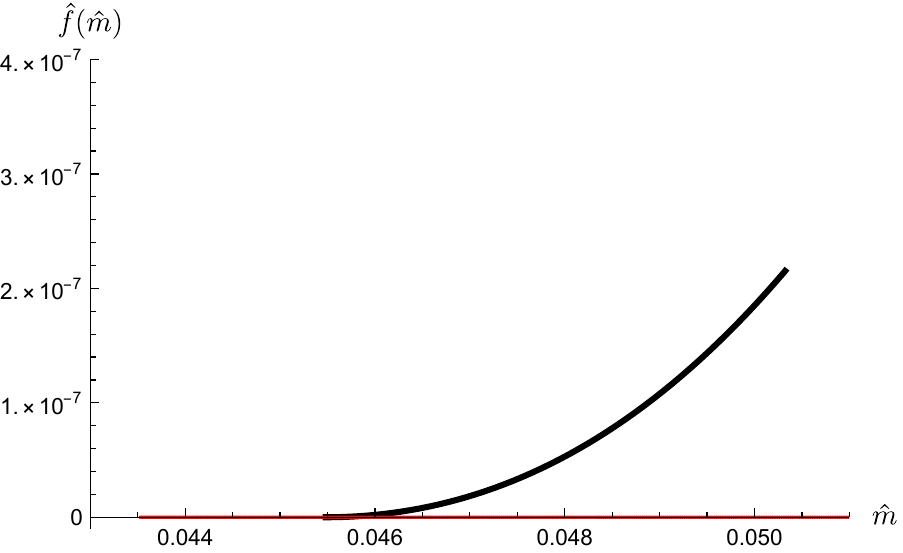}
  \caption{}
  \label{canonicalPSzoom}
\end{subfigure}%
\begin{subfigure}{.32\textwidth}
  \centering
  \includegraphics[width=0.95\linewidth]{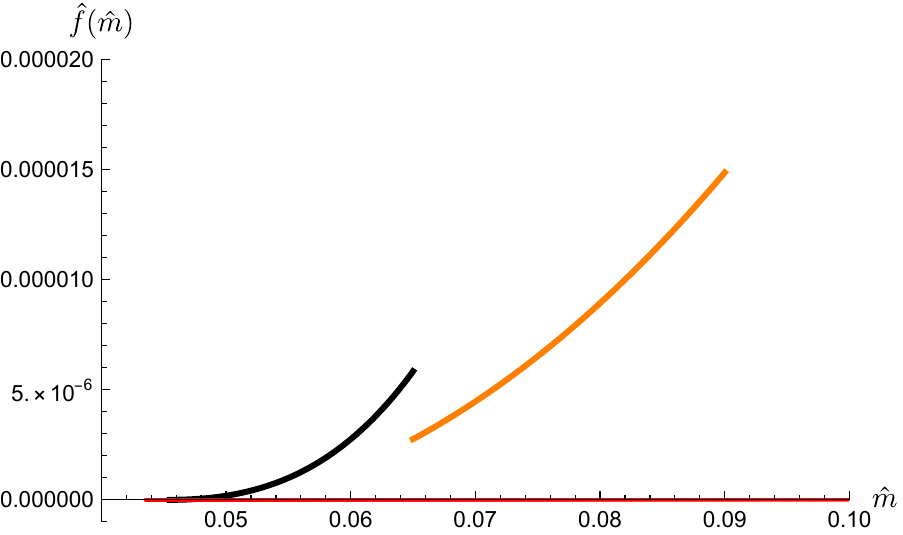}
  \caption{}
  \label{canonicalPSunstable}
\end{subfigure}
\caption{\textit{a) Dimensionless free energy density $\hat{f}$ as a function of $\hat{m}$ for a choice of parameters such that $\frac{\sqrt{N_{3}}}{N_{5}}=45$ and $\frac{N_{3}}{N}=0.1$ for the solution of section \ref{3.0} (black branch) and the numerical solution (red and blue branches) of \cite{Bena_2019}. The rectangle captures the perturbative regime \ref{regime_of_m}.  (b) Zooming in the rectangle of figure (a). There exists a branch of solutions which start life close to the supersymmetric point of \cite{Bena_2019}. (c) Plot of $\hat{f}$ where now we also include the unstable branch of the blackfold solution (orange curve), for the same choice of parameters as in \ref{canonicalPS}, \ref{canonicalPSzoom}}.}
\end{figure}
All quantities of interest are then functions of $\hat{m}=\frac{m}{\sqrt{3}T}$. To obtain $\hat{f}$ for our blackfold solutions we evaluate the free energy density at second second order in the entropy expansion. We also invert locally, around the extremal point, the entropy expansion of the temperature to obtain a relation $S=S(T)$. The constraint of maximum entropy
\begin{equation}\label{constraint_max_S}
    S(T)<S^{*}~~,
\end{equation}
implies a lower bound on the temperature or, equivalently, a maximum bound $\hat{m}_{1}\approx 2\sqrt{\frac{\pi}{3}}\frac{N_{5}}{(1-\eta)\sqrt{N_{3}}}$. Recall that the maximum possible temperature in the stable branch is the extremal one, so that
\begin{equation}\label{regime_of_m}
    (1-\eta)\hat{m}_{1}<\hat{m}<\hat{m}_{1}~~.
\end{equation}
In fig. \ref{canonicalPS}, \ref{canonicalPSzoom} we plot $\hat{f}$ for the (metastable) five-brane solution on top of mass-perturbed $AdS_{5}\times S^{5}$ obtained in section \ref{Non_extremal_PS}, as well as the numerical solution for the mass-deformed thermal D3-brane.

We observe that one of the branches of the numerical solution, the one with non-negative free energy (red branch), approaches in the supersymmetric limit the blackfold solution (black 'ringoid') produced from the blackening of Polchinski-Strassler polarized branes. One can find regimes of parameter space for which the two solutions meet at high temperatures. These black 'ringoids' take generically positive values of the free energy and never dominate the canonical ensemble.

Our final comment concerns the unstable branch of the blackfold solution. In contrast with the metastable branch, here the extremal temperature is the minimum possible temperature that these solutions possess, leading to an upper bound for $\hat{m}$. Analogously, the constraint \eqref{constraint_max_S} that is common for both branches leads to a lower bound for $\hat{m}$. In fig. \ref{canonicalPSunstable} we plot $\hat{f}$ taking into account both branches of the blackfold solution. The behaviour that is shown here appears to be similar to the behaviour of the red branch of the numerical solution in \ref{canonicalPS}. In particular the red branch of the numerical solution can be described as two curves meeting at a cusp with one curve approaching the supersymmetric point and the other curve connecting the red and blue branches. Fig. \ref{canonicalPSunstable} suggests the existence of such "cusp" joining the black and orange curves of the blackfold solution. A higher-order analysis of the blackfold construction will lead to a more precise picture of how these two branches are associated.

\section{Discussion}\label{5.0}
We used the long-wavelength effective theory of black branes to investigate the Polchinski-Strassler mechanism for constructing holographic duals of the $\mathcal{N}=1$* theory. Our aim was two-fold: on the one hand, to take the first steps towards constructing directly in 10 dimensions explicit backreacted supergravity solutions involving polarized D3's into NS5/D5 branes living in a mass-deformed $AdS_{5}\times S^{5}$ geometry, in an appropriate regime of parameter space; and on the other hand, to provide a non-extremal description of the polarized branes of \cite{Polchinski:2000uf} and analytically compute the leading corrections to the extremal configurations due to a thermally excited five-brane worldvolume. To this end, we considered wrapped five-branes with D3 charge moving in the zero temperature mass-perturbed background of \cite{Polchinski:2000uf} as well as in the high temperature deconfined vacuum described by an asymptotically $AdS_{5}\times S^{5}$ black hole in the infrared \cite{Friedman,Bena_2019}.

The results presented in this paper provide strong evidence for the existence of metastable PS polarized branes. By letting entropy and temperature flow into the D3-NS5 system, for both the low and high temperature phases of the theory considered here, we observed a pattern where the metastable black five-brane ceases to exist via a merger with an unstable "thin" state. Both states have horizon topology $\mathbb{R}^3\times \mathbb{S}^2\times \mathbb{S}^3$. For the low temperature phase of section \ref{3.0} the merger occurs at a critical value of the entropy, whose leading expression we determined in a small horizon area expansion. An interesting new feature arising from the analysis of the high temperature phase in section \ref{4.0} is that this merger point can be reached either via a growing horizon of the brane or by heating up the background black hole until a critical temperature. 

A key point in the construction of ref. \cite{Polchinski:2000uf} is the resolution of the infrared singularity - an anticipated feature for flows to non-conformal theories in the infrared, via a brane configuration. Both the normalizable and non-normalizable modes related to relevant operators of conformal dimension $Δ=3$ grow as one flows down to the AdS throat. It was realized in \cite{Polchinski:2000uf} that the coupling of the D3 branes to the non-normalizable modes is precisely what is needed to trigger a polarization process whose outcome - a charged five-brane - is placed at a non-zero AdS radius. In that way the singularity is replaced by an ordinary object, an expanded brane source. Still, one in principle has to account for divergences one is usually confronted with in field theories with defects. When judging the fate of a naked singularity an important issue is whether it can be cloaked behind an horizon \cite{Gubser:2000nd}. For the polarized branes of Polchinski and Strassler this question has remained open and our results imply that the answer is positive.

We conclude with a summary of open questions and future research directions. Given the leading order analysis presented here, a natural next step is to obtain the first order backreacted solution by solving the remaining field equations. In light of the similar calculation of \cite{Nguyen:2021srl} in the context of Klebanov-Strassler, we are confident that such a metric can be obtained. Armed with the explicit first order solution for all the fields one may "hang" probe strings from the AdS boundary stretching into the mass-deformed geometry and check field theory expectations regarding the confinement or screening of quarks. The explicit solution can also be used to determine higher derivative terms to the action \eqref{action}. Note that in the present work we have encountered another example of the DBI/gravity correspondence, a low energy manifestation of the open/closed string duality of ref. \cite{Polchinski_1995}, which has been shown to hold for N coincident D3-branes in flat space \cite{Grignani_2016} and can be derived in the context of the generalised open/closed duality proposed in \cite{Niarchos_2016}. Thus, generalizing the action \eqref{action} by incorporating new corrections can shine light on this duality for D3-NS5 branes.

Another open question concerns the nature of the merger point we face in both the high and low temperature phases considered here. We presented data showing that this transition point can be characterized as a thin-thin merger between 2 D3-NS5 black holes. Since both states have the same horizon topology, this can be at most a geometric transition. We again note that this transition is qualitatively different from the one observed for the anti-D3's at the tip of the Klebanov-Strassler throat \cite{PhysRevLett.122.181601}, while it seems to share certain features with a particular thermal transition (regime III) of the metastable M2-M5 branes in the CGLP background \cite{Armas_2018}. It would be interesting to understand this mechanism better.
 
The results presented here shed new light on the nature of the phase diagram of the $\mathcal{N}=1$* theory. We have found black ringoid solutions with horizon topology $\mathbb{R}^3\times\mathbb{S}^2\times \mathbb{S}^3$ which come from the blackening of the PS polarized branes. We emphasize that these solutions can survive way above the temperature of the (presumed) Hawking-Page transition\footnote{An analytic derivation of the Hawking-Page transition for the $\mathcal{N}=1$* is still missing in the literature. Towards this direction, most progress has been done in \cite{taylorrobinson2001anomalies}, where the solution with an enhanced $SO(3)\times SO(3)$ symmetry was perturbatively constructed and a complete renormalization procedure of the type IIB on-shell action was performed. However, to find such a solution one must give masses to all 4 Weyl fermions, and then the resulting theory is no longer supersymmetric.} separating the theory into confining and deconfining phases. In \cite{Bena_2019} evidence was presented for the appearance of an instability of black holes with $\mathbb{R}^3\times \mathbb{S}^5$ topology whose endpoint is potentially a black ringoid of the kind studied in this paper. Thus these 2 trajectories of the phase space could meet via a topology change transition $\mathbb{R}^3\times \mathbb{S}^5 \to \mathbb{R}^3\times\mathbb{S}^2\times \mathbb{S}^3$. In this paper we showed the existence of such $\mathbb{R}^3\times\mathbb{S}^2\times \mathbb{S}^3$ solutions and in section \ref{4.4} we give evidence for the picture suggested in \cite{Bena_2019}. However, due to the fact that we are performing a perturbative analysis, our solutions are only valid for $\hat m\ll 1$ and as such we were not able to ascertain whether the phase transition found in \cite{Bena_2019} for $\hat m\sim2.15$ is a confinement/deconfinement or a deconfinement/deconfinement phase transition. Our results in section \ref{4.4} shows that our solutions with topology $\mathbb{R}^3\times\mathbb{S}^2\times \mathbb{S}^3$ are not the dominant configuration in the canonical ensemble for $\hat m\ll 1$. This conclusion does not include the possibility of having a black ringoid in the background of a mass-deformed $AdS_{5}\times S^{5}$ black hole, since at higher-orders in the perturbative scheme it may be possible to achieve thermal equilibrium for these configurations. In addition, we considered phase diagrams in an ensemble for which the global entropy is kept fixed, where we observed that the preferable configuration depends on the particular values of the five-brane charges.

It would be very interesting to further elaborate on the connection between various types of solutions and gain a complete view of the phase space of the theory. An interesting way of gaining such additional knowledge is to perform a stability analysis of the configurations obtained here. This can be done by studying the linear spectrum of perturbations around the equilibrium states discussed in section \ref{2.0}. A similar analysis was carried out for black rings \cite{Armas:2019iqs} and D3-NS5 branes in Klebanov-Strassler \cite{Nguyen:2019syc}. Our expectation is that a Gregory-Laflamme-type of instability will be present for certain values of the charges but that there will be a relatively large regime of stability. 

Finally, one could consider applying the blackfold method to find a non-extremal generalisation of the GPPZ solution by wrapping smeared D3-NS5 branes along a compact direction and obtain the GPPZ solution in the extremal limit. This would provide additional understanding of the relation between the GPPZ solution and the vacua described by Polchinski and Strassler \cite{Polchinski:2000uf}. A similar analysis could be performed in analogous setups in M-theory, in particular in the context of polarised M2- \cite{Bena:2000zb} and M5-branes \cite{Bena:2001aw}.

\subsection*{Acknowledgements}
We would like to thank Troels Harmark for useful discussions and to Jorge E. Santos and Oscar J.C. Dias for sharing the data of their work \cite{Bena_2019}. We also would like to thank Vasilis Niarchos, Thomas Van Riet, Matteo Bertolini, Nikolay Bobev, Fridrik Gautason and Jesse van Muiden for useful comments to an earlier draft of this manuscript. In particular we thank Matteo Bertolini for making us aware of this problem. JA is partly supported by the Dutch Institute for Emergent Phenomena (DIEP) cluster at the University of Amsterdam. This work is part of the Delta ITP consortium, a program of the Netherlands Organisation for Scientific Research (NWO) that is funded by the Dutch Ministry of Education, Culture and Science (OCW).

\newpage


\newpage

\providecommand{\href}[2]{#2}\begingroup\raggedright\endgroup



\begin{thebibliography}{10}

\bibitem{maldacena1999large}
J.~Maldacena, \emph{The large-$n$ limit of superconformal field theories and
  supergravity}, {\emph{International journal of theoretical physics}
  {\bfseries 38} (1999) 1113--1133}.

\bibitem{2000}
O.~Aharony, S.~S. Gubser, J.~Maldacena, H.~Ooguri and Y.~Oz, \emph{Large $n$
  field theories, string theory and gravity},
  \href{http://dx.doi.org/10.1016/s0370-1573(99)00083-6}{\emph{Physics Reports}
  {\bfseries 323} (Jan, 2000) 183–386}.

\bibitem{witten1998anti1}
E.~Witten, \emph{Anti de sitter space and holography}, {\emph{arXiv preprint
  hep-th/9802150} (1998) }.

\bibitem{witten1998anti2}
E.~Witten, \emph{Anti-de sitter space, thermal phase transition, and
  confinement in gauge theories}, {\emph{arXiv preprint hep-th/9803131} (1998)
  }.

\bibitem{Donagi_1996}
R.~Donagi and E.~Witten, \emph{Supersymmetric yang-mills theory and integrable
  systems}, \href{http://dx.doi.org/10.1016/0550-3213(95)00609-5}{\emph{Nuclear
  Physics B} {\bfseries 460} (feb, 1996) 299--334}.

\bibitem{Polchinski:2000uf}
J.~Polchinski and M.~J. Strassler, \emph{{The String dual of a confining
  four-dimensional gauge theory}},
  \href{https://arxiv.org/abs/hep-th/0003136}{{\ttfamily hep-th/0003136}}.

\bibitem{Dielectic-branes}
R.~C. Myers, \emph{Dielectric-branes},
  \href{http://dx.doi.org/10.1088/1126-6708/1999/12/022}{\emph{Journal of High
  Energy Physics} {\bfseries 1999} (Dec, 1999) 022–022}.

\bibitem{2000_GPPZ}
L.~Girardello, M.~Petrini, M.~Porrati and A.~Zaffaroni, \emph{The supergravity
  dual of $\mathcal{N}=1$ super yang–mills theory},
  \href{http://dx.doi.org/10.1016/s0550-3213(99)00764-6}{\emph{Nuclear Physics
  B} {\bfseries 569} (Mar, 2000) 451–469}.

\bibitem{Petrini:2018pjk}
M.~Petrini, H.~Samtleben, S.~Schmidt and K.~Skenderis, \emph{{The 10d Uplift of
  the GPPZ Solution}},
  \href{http://dx.doi.org/10.1007/JHEP07(2018)026}{\emph{JHEP} {\bfseries 07}
  (2018) 026}, [\href{https://arxiv.org/abs/1805.01919}{{\ttfamily
  1805.01919}}].

\bibitem{2018_uplift_GPPZ}
N.~Bobev, F.~F. Gautason, B.~E. Niehoff and J.~van Muiden, \emph{Uplifting
  gppz: a ten-dimensional dual of $\mathcal{N}={1}^{*}$},
  \href{http://dx.doi.org/10.1007/jhep10(2018)058}{\emph{Journal of High Energy
  Physics} {\bfseries 2018} (Oct, 2018) }.

\bibitem{2019}
N.~Bobev, F.~F. Gautason, B.~E. Niehoff and J.~van Muiden, \emph{A holographic
  kaleidoscope for $\mathcal{N} = 1^*$},
  \href{http://dx.doi.org/10.1007/jhep10(2019)185}{\emph{Journal of High Energy
  Physics} {\bfseries 2019} (Oct, 2019) }.

\bibitem{Bena_2019}
I.~Bena, O.~J. Dias, G.~S. Hartnett, B.~E. Niehoff and J.~E. Santos,
  \emph{Holographic dual of hot Polchinski-Strassler quark-gluon plasma},
  \href{http://dx.doi.org/10.1007/jhep09(2019)033}{\emph{Journal of High Energy
  Physics} {\bfseries 2019} (Sep, 2019) }.

\bibitem{Friedman}
D.~Z. Freedman and J.~A. Minahan, \emph{Finite temperature effects in the
  supergravity dual of the $\mathcal{N} = 1^*$ gauge theory},
  \href{http://dx.doi.org/10.1088/1126-6708/2001/01/036}{\emph{Journal of High
  Energy Physics} {\bfseries 2001} (Jan, 2001) 036–036}.

\bibitem{Emparan_2010}
R.~Emparan, T.~Harmark, V.~Niarchose and N.~A. Obers, \emph{Essentials of
  blackfold dynamics},
  \href{http://dx.doi.org/10.1007/jhep03(2010)063}{\emph{Journal of High Energy
  Physics} {\bfseries 2010} (mar, 2010) }.

\bibitem{Blackfolds_string_theory}
R.~Emparan, T.~Harmark, V.~Niarchos and N.~A. Obers, \emph{Blackfolds in
  supergravity and string theory},
  \href{http://dx.doi.org/10.1007/jhep08(2011)154}{\emph{Journal of High Energy
  Physics} {\bfseries 2011} (aug, 2011) }.

\bibitem{PhysRevLett.122.181601}
J.~Armas, N.~Nguyen, V.~Niarchos, N.~A. Obers and T.~Van~Riet, \emph{Metastable
  nonextremal antibranes},
  \href{http://dx.doi.org/10.1103/PhysRevLett.122.181601}{\emph{Phys. Rev.
  Lett.} {\bfseries 122} (May, 2019) 181601}.

\bibitem{m2-m52019}
J.~Armas, N.~Nguyen, V.~Niarchos and N.~A. Obers, \emph{Thermal transitions of
  metastable m-branes},
  \href{http://dx.doi.org/10.1007/jhep08(2019)128}{\emph{Journal of High Energy
  Physics} {\bfseries 2019} (Aug, 2019) }.

\bibitem{2012a}
J.~Armas, T.~Harmark, N.~A. Obers, M.~Orselli and A.~V. Pedersen, \emph{Thermal
  giant gravitons},
  \href{http://dx.doi.org/10.1007/jhep11(2012)123}{\emph{Journal of High Energy
  Physics} {\bfseries 2012} (Nov, 2012) }.

\bibitem{Armas:2013ota}
J.~Armas, N.~A. Obers and A.~V. Pedersen, \emph{{Null-Wave Giant Gravitons from
  Thermal Spinning Brane Probes}},
  \href{http://dx.doi.org/10.1007/JHEP10(2013)109}{\emph{JHEP} {\bfseries 10}
  (2013) 109}, [\href{https://arxiv.org/abs/1306.2633}{{\ttfamily 1306.2633}}].

\bibitem{2012b}
G.~Grignani, T.~Harmark, A.~Marini, N.~A. Obers and M.~Orselli, \emph{Thermal
  string probes in ads and finite temperature wilson loops},
  \href{http://dx.doi.org/10.1007/jhep06(2012)144}{\emph{Journal of High Energy
  Physics} {\bfseries 2012} (Jun, 2012) }.

\bibitem{Armas:2014nea}
J.~Armas and M.~Blau, \emph{{Black probes of Schr\"odinger spacetimes}},
  \href{http://dx.doi.org/10.1007/JHEP08(2014)140}{\emph{JHEP} {\bfseries 08}
  (2014) 140}, [\href{https://arxiv.org/abs/1405.1301}{{\ttfamily 1405.1301}}].

\bibitem{Armas:2010hz}
J.~Armas and N.~A. Obers, \emph{{Blackfolds in (Anti)-de Sitter Backgrounds}},
  \href{http://dx.doi.org/10.1103/PhysRevD.83.084039}{\emph{Phys. Rev. D}
  {\bfseries 83} (2011) 084039},
  [\href{https://arxiv.org/abs/1012.5081}{{\ttfamily 1012.5081}}].

\bibitem{Armas:2015qsv}
J.~Armas, N.~A. Obers and M.~Sanchioni, \emph{{Gravitational Tension, Spacetime
  Pressure and Black Hole Volume}},
  \href{http://dx.doi.org/10.1007/JHEP09(2016)124}{\emph{JHEP} {\bfseries 09}
  (2016) 124}, [\href{https://arxiv.org/abs/1512.09106}{{\ttfamily
  1512.09106}}].

\bibitem{2016}
J.~Armas, J.~Gath, V.~Niarchos, N.~A. Obers and A.~V. Pedersen, \emph{Forced
  fluid dynamics from blackfolds in general supergravity backgrounds},
  \href{http://dx.doi.org/10.1007/jhep10(2016)154}{\emph{Journal of High Energy
  Physics} {\bfseries 2016} (Oct, 2016) }.

\bibitem{Camps_2012}
J.~Camps and R.~Emparan, \emph{Derivation of the blackfold effective theory},
  \href{http://dx.doi.org/10.1007/jhep03(2012)038}{\emph{Journal of High Energy
  Physics} {\bfseries 2012} (mar, 2012) }.

\bibitem{Niarchos_2016}
V.~Niarchos, \emph{Open/closed string duality and relativistic fluids},
  \href{http://dx.doi.org/10.1103/physrevd.94.026009}{\emph{Physical Review D}
  {\bfseries 94} (jul, 2016) }.

\bibitem{Nguyen:2021srl}
N.~Nguyen and V.~Niarchos, \emph{{On matched asymptotic expansions of
  backreacting metastable anti-branes}},
  \href{http://dx.doi.org/10.1007/JHEP06(2022)055}{\emph{JHEP} {\bfseries 06}
  (2022) 055}, [\href{https://arxiv.org/abs/2112.04514}{{\ttfamily
  2112.04514}}].

\bibitem{https://doi.org/10.48550/arxiv.hep-th/0006117}
D.~Marolf, \emph{Chern-simons terms and the three notions of charge},  2000.
\newblock 10.48550/ARXIV.HEP-TH/0006117.

\bibitem{Gaiotto_2015}
D.~Gaiotto, A.~Kapustin, N.~Seiberg and B.~Willett, \emph{Generalized global
  symmetries}, \href{http://dx.doi.org/10.1007/jhep02(2015)172}{\emph{Journal
  of High Energy Physics} {\bfseries 2015} (feb, 2015) }.

\bibitem{Armas_2018}
J.~Armas, J.~Gath, A.~Jain and A.~V. Pedersen, \emph{Dissipative hydrodynamics
  with higher-form symmetry},
  \href{http://dx.doi.org/10.1007/jhep05(2018)192}{\emph{Journal of High Energy
  Physics} {\bfseries 2018} (may, 2018) }.

\bibitem{https://doi.org/10.48550/arxiv.2202.04655}
S.~Kaya and T.~Rudelius, \emph{Higher-group symmetries and weak gravity
  conjecture mixing},  2022.
\newblock 10.48550/ARXIV.2202.04655.

\bibitem{Caldarelli_2009}
M.~M. Caldarelli, {\'{O}}.~J. Dias, R.~Emparan and D.~Klemm, \emph{Black holes
  as lumps of fluid},
  \href{http://dx.doi.org/10.1088/1126-6708/2009/04/024}{\emph{Journal of High
  Energy Physics} {\bfseries 2009} (apr, 2009) 024--024}.

\bibitem{Cohen_Maldonado_2016}
D.~Cohen-Maldonado, J.~Diaz, T.~V. Riet and B.~Vercnocke, \emph{Observations on
  fluxes near anti-branes},
  \href{http://dx.doi.org/10.1007/jhep01(2016)126}{\emph{Journal of High Energy
  Physics} {\bfseries 2016} (jan, 2016) }.

\bibitem{Vafa_1994}
C.~Vafa and E.~Witten, \emph{A strong coupling test of s-duality},
  \href{http://dx.doi.org/10.1016/0550-3213(94)90097-3}{\emph{Nuclear Physics
  B} {\bfseries 431} (dec, 1994) 3--77}.

\bibitem{Dorey_1999}
N.~Dorey, \emph{An elliptic superpotential for softly broken $\mathcal{N}=4$
  supersymmetric yang-mills theory},
  \href{http://dx.doi.org/10.1088/1126-6708/1999/07/021}{\emph{Journal of High
  Energy Physics} {\bfseries 1999} (jul, 1999) 021--021}.

\bibitem{Kinar_2001}
Y.~Kinar, A.~Loewy, E.~Schreiber, J.~Sonnenschein and S.~Yankielowicz,
  \emph{Supergravity and worldvolume physics in the dual description of
  $\mathcal{N} = 1^{\star}$ theory},
  \href{http://dx.doi.org/10.1088/1126-6708/2001/03/013}{\emph{Journal of High
  Energy Physics} {\bfseries 2001} (mar, 2001) 013--013}.

\bibitem{Emparan:2003sy}
R.~Emparan and R.~C. Myers, \emph{{Instability of ultra-spinning black holes}},
  \href{http://dx.doi.org/10.1088/1126-6708/2003/09/025}{\emph{JHEP} {\bfseries
  09} (2003) 025}, [\href{https://arxiv.org/abs/hep-th/0308056}{{\ttfamily
  hep-th/0308056}}].

\bibitem{Emparan:2009vd}
R.~Emparan, T.~Harmark, V.~Niarchos and N.~A. Obers, \emph{{New Horizons for
  Black Holes and Branes}},
  \href{http://dx.doi.org/10.1007/JHEP04(2010)046}{\emph{JHEP} {\bfseries 04}
  (2010) 046}, [\href{https://arxiv.org/abs/0912.2352}{{\ttfamily 0912.2352}}].

\bibitem{Dias:2019wof}
O.~J.~C. Dias, G.~S. Hartnett and J.~E. Santos, \emph{{Covariant Noether
  charges for type IIB and 11-dimensional supergravities}},
  \href{http://dx.doi.org/10.1088/1361-6382/abc136}{\emph{Class. Quant. Grav.}
  {\bfseries 38} (2021) 015003},
  [\href{https://arxiv.org/abs/1912.01030}{{\ttfamily 1912.01030}}].

\bibitem{Gubser:2000nd}
S.~S. Gubser, \emph{{Curvature singularities: The Good, the bad, and the
  naked}}, \href{http://dx.doi.org/10.4310/ATMP.2000.v4.n3.a6}{\emph{Adv.
  Theor. Math. Phys.} {\bfseries 4} (2000) 679--745},
  [\href{https://arxiv.org/abs/hep-th/0002160}{{\ttfamily hep-th/0002160}}].

\bibitem{Polchinski_1995}
J.~Polchinski, \emph{Dirichlet branes and ramond-ramond charges},
  \href{http://dx.doi.org/10.1103/physrevlett.75.4724}{\emph{Physical Review
  Letters} {\bfseries 75} (dec, 1995) 4724--4727}.

\bibitem{Grignani_2016}
G.~Grignani, T.~Harmark, A.~Marini and M.~Orselli, \emph{The Born-Infeld/Gravity
  correspondence},
  \href{http://dx.doi.org/10.1103/physrevd.94.066009}{\emph{Physical Review D}
  {\bfseries 94} (sep, 2016) }.

\bibitem{taylorrobinson2001anomalies}
M.~Taylor-Robinson, \emph{Anomalies, counterterms and the $\mathcal {N} =0$
  Polchinski-Strassler solutions},  2001.

\bibitem{Armas:2019iqs}
J.~Armas and E.~Parisini, \emph{{Instabilities of Thin Black Rings: Closing the
  Gap}}, \href{http://dx.doi.org/10.1007/JHEP04(2019)169}{\emph{JHEP}
  {\bfseries 04} (2019) 169},
  [\href{https://arxiv.org/abs/1901.09369}{{\ttfamily 1901.09369}}].

\bibitem{Nguyen:2019syc}
N.~Nguyen, \emph{{Comments on the stability of the KPV state}},
  \href{http://dx.doi.org/10.1007/JHEP11(2020)055}{\emph{JHEP} {\bfseries 11}
  (2020) 055}, [\href{https://arxiv.org/abs/1912.04646}{{\ttfamily
  1912.04646}}].

\bibitem{Bena:2000zb}
I.~Bena, \emph{{The M theory dual of a three-dimensional theory with reduced
  supersymmetry}},
  \href{http://dx.doi.org/10.1103/PhysRevD.62.126006}{\emph{Phys. Rev. D}
  {\bfseries 62} (2000) 126006},
  [\href{https://arxiv.org/abs/hep-th/0004142}{{\ttfamily hep-th/0004142}}].

\bibitem{Bena:2001aw}
I.~Bena and D.~Vaman, \emph{{The Polarization of M5 branes and little string
  theories with reduced supersymmetry}},
  \href{http://dx.doi.org/10.1088/1126-6708/2001/11/032}{\emph{JHEP} {\bfseries
  11} (2001) 032}, [\href{https://arxiv.org/abs/hep-th/0101064}{{\ttfamily
  hep-th/0101064}}].

\end{thebibliography}
\end{document}